%% file: 0_ijcai26.tex
\documentclass{article}
\usepackage{ijcai26}

\usepackage{times}
\usepackage{soul}
\usepackage{url}
\usepackage[hidelinks]{hyperref}
\usepackage[utf8]{inputenc}
\usepackage[small]{caption}
\usepackage{graphicx}
\usepackage{amsmath}
\usepackage{amsthm}
\usepackage{booktabs}
\usepackage[switch]{lineno}

\usepackage{bm}
\usepackage{amssymb}
\usepackage{multirow}
\usepackage{array}
\usepackage{ascmac}
\usepackage{bbm}
\usepackage{subcaption}
\usepackage{appendix}
\usepackage[export]{adjustbox}
\usepackage{xcolor}
\usepackage{lipsum}
\newcommand\blfootnote[1]{%
  \begingroup
  \renewcommand\thefootnote{}\footnote{#1}%
  \addtocounter{footnote}{-1}%
  \endgroup
}
\usepackage{longtable}
\usepackage{placeins}
\usepackage{needspace}
\usepackage{float}
\usepackage{algorithm}      
\usepackage{algpseudocode}  

\title{Psychological Benefits and Costs of Diversifying Algorithmic Recourse}

\author{
Tomu Tominaga$^1$
\and
Naomi Yamashita$^2$
\and
Takeshi Kurashima$^1$\\
\affiliations
$^1$NTT Human Informatics Laboratories, NTT, Inc.\\
$^2$Graduate School of Informatics, Kyoto University\\
\emails
tomu.tominaga@ntt.com,
naomiy@acm.org,
takeshi.kurashima@ntt.com
}

\begin{document}

\maketitle

\begin{abstract}
Algorithmic recourse provides counterfactual action plans that help people overturn unfavorable AI decisions. 
While diverse recourse sets may improve transparency and motivation, they may also impose cognitive load and negative emotions by increasing counterfactual reasoning demands. 
To examine this trade-off, we conducted a between-subjects controlled experiment ($N{=}750$) that manipulated recourse-set diversity and size, and evaluated these effects on psychological benefits and costs. 
Results show that diversification enhances psychological benefits (e.g., willingness to act) for small sets without incurring additional psychological costs, whereas for large sets, it makes cognitive load more salient. 
These findings suggest that naively diversifying recourse can burden decision subjects, underscoring the need for new diversification methods that incorporate human cognition and psychology to mitigate such costs.
\blfootnote{This paper is an extended version of our paper accepted at IJCAI-ECAI 2026, including supplementary materials.}

\end{abstract}

\input{1_introduction}
\input{2_related_work}
\input{3_method}
\input{4_results}
\input{5_discussion}
\input{6_conclusion}

\section*{Ethical Statement}
This study involving human subjects was reviewed and approved by the external ethics review board, the Institutional Review Board of Public Health Research Foundation (approval ID: PHRF-IRB 25A0004), and all procedures were conducted in accordance with the guidelines.

\bibliographystyle{named}
\bibliography{ijcai26}

\input{7_supplementary}

\end{document}

%% file: 1_introduction.tex
\section{Introduction}
As AI models are increasingly being deployed in high-stakes domains, algorithmic recourse has become a central topic in explainable AI as a way to address the ``right to explanation''~\cite{Wachter2018}.
Recourse takes the form of counterfactual explanations that suggest actionable changes allowing decision subjects to flip adverse decisions~\cite{Ustun2019,Karimi2021}.
For example, a loan applicant denied by an AI system might be told, ``If your annual income were \$1,000 higher, you would be approved.''  
Ideally, such recourse should be both understandable as an explanation and actionable as a behavioral plan~\cite{Karimi2022}.

To this end, a promising approach is to diversify recourse options by generating multiple counterfactuals that are mutually dissimilar in the feature space~\cite{Mothilal2020}.
Prior work suggests that presenting diverse recourse options can help people grasp complex concepts~\cite{Wang2019,Spiro1989}, increase the chance that at least one plan is feasible under real-world constraints~\cite{Barocas2020}, and strengthen autonomy and motivation by letting decision subjects actively choose among alternatives~\cite{Ryan2000}.
However, counterfactual reasoning often imposes cognitive demands and emotional burdens because it forces individuals to consider negative and positive outcomes simultaneously~\cite{Byrne2005,Byrne2007} or recall alternative pasts and personal shortcoming~\cite{Byrne2016,Tominaga2025}.
Although the trade-off between psychological benefits and costs is theoretically anticipated, it has rarely been examined empirically.  
Consequently, it remains unclear how recourse should be diversified in practice.

\begin{figure}
    \centering
    \includegraphics[width=0.8\columnwidth]{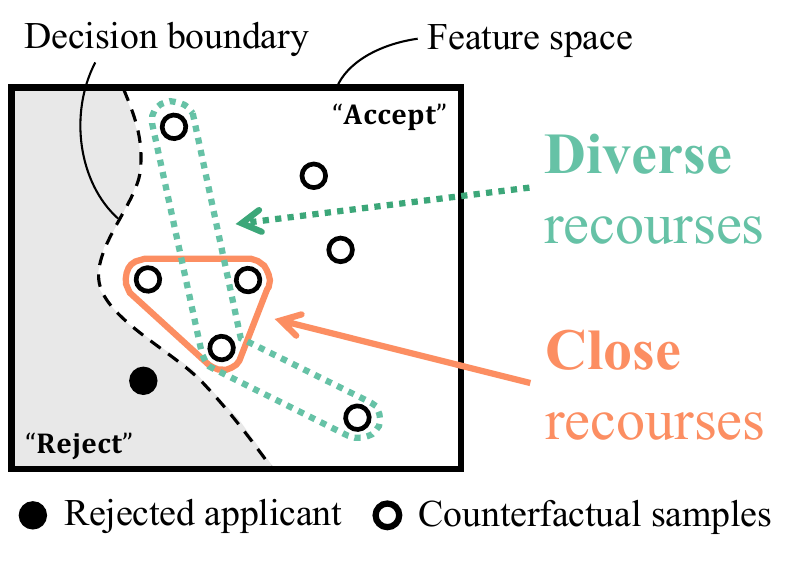}
    \caption{Conceptual difference in counterfactual sample selection between ``Close'' and ``Diverse'' conditions, illustrated with an example of three options. While the ``Close'' condition only focuses on proximity of counterfactuals, the ``Diverse'' condition focuses on both proxmity and mutual dissimilarity of counterfactuals.}
    \label{fig:concept}
\end{figure}

To address this gap, we conducted a between-subjects experiment ($N{=}750$) in the context of automobile loan applications, manipulating both the number and diversity of recourse options.
We generated sets of size 1, 3, or 7 using two methods: (1) selecting counterfactuals closest to the participant's data (``Close'' condition), or (2) selecting counterfactuals close to the data but dissimilar from one another (``Diverse'' condition).  
Figure~\ref{fig:concept} illustrates the conceptual difference between these two approaches.
Participants evaluated the recourse set with regard to subjective reasonability, subjective actionability, willingness to act, and decision acceptance as indicators of benefits, and cognitive load and emotional burdens as measures of costs.
This design allowed us to address the following research question (RQ): \textit{\textbf{How do recourse-set diversity and size affect decision subjects' psychological benefits and costs?}}

Addressing this RQ, we found contrasting patterns for benefits and costs.
Diverse sets improved benefits from one to three options but then plateaued, while Close sets demonstrated stepwise gains from one to three to seven, matching Diverse sets only at seven.
This trend was particularly pronounced in willingness to act.
In contrast, cognitive load remained for Close sets beyond three, whereas it increased with set size for Diverse sets and became most salient at seven.
Open-ended responses echoed this pattern; diverse sets helped participants find a meaningful plan at three options, but at seven they more often questioned the clarity of the AI's decision criteria.
Taken together, these results suggest that diversification is promising for relatively small sets, but calls for safeguards when presenting larger sets.

This study makes three contributions: (1) we provide empirical evidence on how recourse diversity shapes participants' perceptions and experiences, bridging the gap between theoretical expectations and practice, (2) we identify the conditions under which the balance between benefits and costs is optimized, showing that diversification with relatively few options enhances motivation while minimizing psychological strain, and (3) based on these insights, we offer design implications to maximize the use of diversification, and discuss future directions for establishing such methods.

%% file: 2_related_work.tex
\section{Related Work\label{sec:related_work}}
To aid decision subjects who receive adverse AI judgments by clarifying its reasons and suggesting actionable steps, most research frames recourse generation as identifying the nearest counterfactual sample, an individual with a positive outcome whose profile is close to the target user~\cite{Wachter2018}.
Formally, given an $N$-dimensional feature space $\mathcal{X} = \mathcal{X}_1 \times ... \times \mathcal{X}_N$, a binary classifier $f: \mathcal{X} \rightarrow \{+1,-1\}$, a distance function $\mathrm{dist}: \mathcal{X} \times \mathcal{X} \rightarrow \mathbb{R}_{\geq 0}$, the task is to find the closest $\bm{c^*}$ in the positive region $\mathcal{A}=\{\bm{c} \in \mathcal{X} \mid f(\bm{c})\neq f(\bm{x})\}$ for a target user $\bm{x} \in \mathcal{X}$ with $f(\bm{x})=-1$:
\begin{align}\label{eq:computing_recourse_sample}
    \bm{c^*} &\in \underset{\bm{c}\in \mathcal{A}}{\operatorname{argmin}}\ \mathrm{dist}(\bm{c}, \bm{x}),
\end{align}
and then construct the perturbation $\bm{\delta}=\bm{c^*}-\bm{x}$ as a recourse and provide it to the user.
However, it is challenging to provide decision subjects with recourse both acceptable and actionable~\cite{Tominaga2024}, and researchers have proposed many techniques to tackle this problem (see extensive reviews~\cite{Karimi2022,Verma2024}).

Among them, one promising direction is to diversify recourse options~\cite{Laugel2023}.
This approach generates and presents multiple, varied recourse options so that decision subjects can identify an option that aligns with their circumstances and preferences~\cite{Wachter2018,Barocas2020}.
Prior work proposes computational algorithms that force different recourse options to suggest different actions~\cite{Ustun2019,Russell2019} or introduces the term that maximizes pairwise dissimilarity among counterfactuals in the feature space~\cite{Mothilal2020}.

Diversification is expected to benefit decision subjects in two ways: as explanations and as recommendations.
As explanations, diverse recourse options may help decision subjects infer decision criteria because multifaceted explanations can improve understanding~\cite{Wang2019,Spiro1989}.
Since better understanding is associated with stronger acceptance of unfavorable outcomes~\cite{Tominaga2025}, a diverse set may also make the explanation more persuasive.
As recommendations, diversification can support both actionability and motivation.
Specifically, a broader set of options increases the chance that at least one option is actionable under real-world constraints~\cite{Wachter2018,Barocas2020}.
In addition, comparing alternatives and choosing a best fit can strengthen autonomy, a key source of motivation~\cite{Ryan2000}; therefore, offering diverse recourse options can enhance decision subjects' autonomy and thereby increase their willingness to act.

At the same time, diversification may impose cognitive and emotional costs.
Counterfactual thinking can be more demanding than causal thinking~\cite{Byrne2005,Byrne2007} as it requires people to reason about dual possibilities (e.g., if X' instead of X, then Y' instead of Y)~\cite{McEleney2006}.
It can also elicit negative emotions because contrasting reality with counterfactual alternatives often evokes regret, guilt, or self-blame~\cite{Byrne2016}.
In fact, recent work found that participants sometimes perceived recourse as blaming their past shortcomings~\cite{Tominaga2025}.
Since diversification exposes decision subjects to more varied counterfactuals, these psychological costs may intensify, especially as the number of recourse options increases.

Taken together, while diverse recourse sets are promising, they may also raise concerns.
This trade-off has not yet been empirically examined, leaving open how much diversity should be introduced.
Our study provides empirical evidence on this issue and identifies conditions for optimal diversity control in algorithmic recourse.

%% file: 3_method.tex
\section{Method}
To address our research question, we conducted a randomized between-subjects experiment manipulating the diversity and size of the set of recourse options, and elucidated the effects of these factors on psychological benefits and costs.

\subsection{Experimental Setup\label{subsec:setup}}
\subsubsection{Hypothetical Scenario}
Based on prior studies~\cite{Tominaga2024,Tominaga2025}, we crafted a hypothetical scenario of a car loan screening because credit assessment is a central high-stakes domain in the research areas of machine learning~\cite{Karimi2022,Verma2024} and human-computer interaction~\cite{Wang2023,Gemalmaz2022,Lyons2022}.
Specifically, participants engaged in the following scenario: \textit{Participants were asked to imagine applying for a two-year auto loan equal to one-third of their annual income and submit their profile data to a financial institution that uses AI-based credit assessment. They then received a rejection notice with a recourse set and were instructed to evaluate its content.}
In this scenario, the institution rejects loan applications when the requested amount is at least one-quarter of the applicant's annual income. 
Thus, our experiment simulates a setting where all participants were rejected and would be approved if their profile corresponded to an annual income at least 4/3 times their current income. 
This rule is not disclosed to them.

The repayment period was fixed at two years, and the loan size was defined relative to each participant's income. 
This avoided variation in distance from participants to the decision boundary, which could affect their recourse evaluations.  

\subsubsection{Real Profile Data for Screening}
Under this scenario, participants submitted their real profile data.
The profile data items are shown in Table~\ref{tab:profile}.
\begin{table}[t]
    \setlength{\tabcolsep}{2pt}
    \centering
    \begin{tabular}{clcc}
        \toprule
        \# & Profile item (feature) & Con. & Typ.\\
        \midrule
        1 & Residence (e.g., Tokyo) & -- & $\mathrm{cat}$\\
        2 & Type of residence (e.g., rental) &  -- & $\mathrm{cat}$\\
        3 & Education (e.g., high school) & $\uparrow$ & $\mathrm{ord}$\\
        4 & Employment (e.g., private company)& -- & $\mathrm{cat}$\\
        5 & Job title (e.g., employee)& $\uparrow$ & $\mathrm{ord}$\\
        6 & Years of service (e.g., 1-3 years) & $\uparrow$ & $\mathrm{ord}$\\
        7 & Management career (e.g., 1-3 years) & $\uparrow$ & $\mathrm{ord}$\\
        8 & Daily working hours (e.g., 0-2 hours) & -- & $\mathrm{ord}$\\
        9 & Daily remote working hours (e.g., 0-2 hours) & -- & $\mathrm{ord}$\\
        10 & Number of side jobs (e.g., 2 jobs) & -- & $\mathrm{ord}$\\
        11 & Job change experience (e.g., yes) & $\uparrow$ & $\mathrm{ord}$\\
        12 & Overseas work experience (e.g., yes) & $\uparrow$ & $\mathrm{ord}$\\
        13 & Study abroad experience (e.g., yes) & $\uparrow$ & $\mathrm{ord}$\\
        14 & Best TOEIC$^\dagger$ score (e.g., 600-695) & $\uparrow$ & $\mathrm{ord}$\\
        15 & Facebook use (e.g., yes) & -- & $\mathrm{cat}$\\
        16 & LinkedIn use (e.g., yes) & -- & $\mathrm{cat}$\\
        \bottomrule
    \end{tabular}
    \caption{Profile items used in the scenario. ``Con.'' indicates the allowed direction of change in recourse (\(\uparrow\): increase-only; --: no direction constraint), which defines the feasible/mutable zone \(\mathcal{Z}\) (Equation~(\ref{eq:computing_recourse_sample})). ``Typ.'' denotes feature type: categorical ($\mathrm{cat}$) or ordinal ($\mathrm{ord}$). \(^{\dagger}\)TOEIC is a widely used English proficiency test in Japan.\label{tab:profile}}
\end{table}
Based on prior studies~\cite{Tominaga2024,Tominaga2025}, we adopted 16 items: basic attributes (\#1–\#3), current employment (\#4–\#10), job experience and skills (\#11–\#14), and personal networks (\#15–\#16).  
These features commonly appear in loan studies~\cite{Yurrita2023}, credit evaluation datasets~\cite{Becker1996,Yeh2016,Hoffman2013}, and typical resumes~\cite{Brown1994,Cole2007}.
For the details of answer options and their feature values, see Supplementary Materials (SM), Table~\ref{supp_tab:profile_data}.

Immutable attributes such as gender were not included as they cannot be altered by individual effort~\cite{Kirfel2021}. 
We also excluded participants' income from the recourse attributes.  
In our scenario, all participants were rejected for insufficient income.  
If income were included, all recourses would propose increasing it, sharply limiting diversity.  
To observe user reactions to diverse recourses, we did not use income as a profile item in recourses.

\subsection{Factor Design and Manipulations\label{subsec:manipulation}}
\subsubsection{Factors}\label{subsubsec:factor}
This study employed a two-factor factorial design manipulating recourse-set diversity and size. 
Recourse-set diversity captures the extent to which multiple recourse options are mutually distant in the feature space. 
To isolate the effect of diversity from the number of options, we independently varied set size, enabling a precise test of how diversity shapes participants' reactions.
Each factor was designed as follows.

\paragraph{Diversity: Close or Diverse.}
To control recourse-set diversity, we constructed two types of recourse sets: (1) a \textit{Close} set, formed by selecting counterfactual samples based solely on proximity to the decision subject~\cite{Wachter2018}, and (2) a \textit{Diverse} set, formed by selecting samples based on proximity and pairwise dissimilarity among the selected samples~\cite{Mothilal2020}.

\paragraph{Size: 1, 3, or 7.} 
We varied the number of recourse options as 1, 3, and 7. 
The single-option condition served as a reference condition for evaluating multi-option effects.
We set the minimum set size to three.
This is because, compared with only two options, three or more options enable comparisons across multiple pairs, 
facilitating the grouping of options~\cite{Tversky1977,Medin1993} and thus helping individuals capture the spread and relational structure of options in the sample space, thereby enhancing the perception of diversity~\cite{Kahn2004}.
We set seven as an upper bound, expecting cognitive load to saturate near this range.
Conventional work on short-term memory highlights limits on retaining and comparing ``7$\pm$2'' items~\cite{Miller1956}, and later evidence shows that too many options can be ignored or yield random choices~\cite{Scheibehenne2010}. 

\subsubsection{Recourse Computation to Experimental Conditions\label{subsubsec:recourse_computation}}
In this experiment, we randomly assign participants to one of the experimental conditions defined by two factors: recourse-set diversity $p:=\text{Close},\text{Diverse}$ and size $k:=1,3,7$. 
Hereafter, we denote each condition using the $p$--$k$ notation (e.g., Diverse--3). 
Since diversity is undefined for a singleton set, we exclude the Diverse--1 condition, resulting in five conditions in total (i.e., $2\times3-1$).

To construct counterfactual samples for recourse sets, we used an existing dataset of 4,057 profile entries collected in prior work~\cite{Tominaga2024}.
We treated each record as a candidate profile $\bm{c}\in\mathcal{X}$ and restrict candidates to an admissible area $\mathcal{A}\cap\mathcal{Z}$.
$\mathcal{A}$ denotes the set of distinct profiles that would be approved; in our loan scenario, this corresponds to profiles that satisfy the income-related approval condition (Section~\ref{subsec:setup}).
$\mathcal{Z}$ encodes constraints on the direction of change for each feature.
As shown in Table~\ref{tab:profile}, we imposed these constraints to ensure that recourse suggestions remain plausible (e.g., ``Education'' is allowed to increase only).

Given a participant profile $\bm{x}\in\mathcal{X}$, we select a set of $k$ counterfactual samples $C:=\{\bm{c_1},...,\bm{c_k}\}$ by solving:
\begin{equation}
    \resizebox{\linewidth}{!}{$
        \displaystyle
        C^{*}= \underset{\substack{C \subset (\mathcal{A}\cap\mathcal{Z}) \\ |C|=k}}{\operatorname{argmin}}\ \frac{1}{k}\sum_{\bm{c}\in C}\mathrm{dist}(\bm{c},\bm{x}) - \lambda\cdot \frac{1}{\binom{k}{2}}\sum_{\substack{\bm{c_i},\bm{c_j}\in C \\ i<j}}\mathrm{dist}(\bm{c_i},\bm{c_j}),
    $}
    \label{eq:computing_recourse_set}
\end{equation}
where the first term captures proximity (i.e., average distance to $\bm{x}$) and the second term measures diversity (i.e., average pairwise distance).
We set $\lambda=1$ for Diverse and $\lambda=0$ for Close, so the diversity term is considered only for Diverse.
We then compute the recourse set $\Delta=\{\bm{c}-\bm{x} \mid \bm{c}\in C^{*}\}$.

Inspired by prior studies~\cite{Mothilal2020,Wachter2018}, $\mathrm{dist}$ (Equation~(\ref{eq:computing_recourse_set})) is defined as a feature-wise $\ell_1$ distance between given samples $\bm{u},\bm{v}\in\mathcal{X}$:
\begin{equation}
    \resizebox{\linewidth}{!}{$
        \displaystyle
        \mathrm{dist}(\bm{u},\bm{v})=\frac{1}{|\mathcal{I}_{\mathrm{cat}}|}\sum_{i\in \mathcal{I}_{\mathrm{cat}}}\mathbbm{1}[\bm{u}_i\neq \bm{v}_i] + \frac{1}{|\mathcal{I}_{\mathrm{ord}}|}\sum_{i\in \mathcal{I}_{\mathrm{ord}}}\frac{|\bm{u}_i-\bm{v}_i|}{\mathrm{MAD}_i}.
    $}
    \label{eq:recourse_dist}
\end{equation}
Here, $\mathbbm{1}$ is the indicator function (1 if true, 0 otherwise).
$\mathcal{I}_{\mathrm{cat}}$ and $\mathcal{I}_{\mathrm{ord}}$ are the sets of categorical and ordinal features, respectively; $\bm{u}_i$ and $\bm{v}_i$ denote the option values of feature $i$. 
The first term averages mismatches over categorical features, while the second term averages absolute differences over ordinal features, normalized by the feature-wise median absolute deviation $\mathrm{MAD}_i$.

To solve Equation~(\ref{eq:computing_recourse_set}), we used a greedy approximation (see SM, Section~\ref{supp_sec:alg} for pseudocode) because our goal is stimulus generation with controlled diversity rather than exact optimization.
Importantly, this procedure produced a clear separation in perceived diversity in our manipulation check (Section~\ref{subsec:procedure}), validating its adequacy for the user study.

\subsection{Measured Outcomes\label{subsec:outcome_measures}}
As described in Section~\ref{sec:related_work}, recourse-set diversification possibly brings various psychological benefits and costs to decision subjects.
To capture them, this study used a questionnaire survey (see SM, Section~\ref{supp_sec:survey} for the full instructions).

To improve the validity of our measurements, we included a comprehension task prior to the measurements (see SM, Section~\ref{supp_subsec:comprehension_check}). 
In this task, participants were instructed to identify  the profile items that the recourse options cited as reasons for rejection and recommended actions for future approval. 
Their responses inconsistent with the presented recourse set were excluded from the analyses.
This task screened out inattentive or low-effort responses, and ensured that subsequent ratings were based on an accurate understanding of the recourse.
Psychological benefits and costs were measured after the comprehension task as follows.

\subsubsection{Psychological Benefits}
Assuming that decision subjects who are provided with a recourse set undergo a sequential experience of judging whether the suggested plan is a reasonable
explanation~\cite{Tominaga2024}, assessing whether it is feasible to
carry out~\cite{Kirfel2021,Wang2023}, forming motivation to act~\cite{Tominaga2024}, and integrating these impressions into their ultimate acceptance of the AI
decision~\cite{Tominaga2025}, we captured psychological benefits with the following four dimensions:
\begin{description}
    \item[Subjective reasonability:] the extent to which participants found the recourse option to be a reasonable explanation for the rejection.
    \item[Subjective actionability:] the extent to which participants found the recourse option easy or difficult to carry out.
    \item[Willingness to act:] the extent to which participants were willing to carry out the recourse option.
    \item[Decision acceptance:] the extent to which participants accept the AI decision outcome of their loan application.
\end{description}
All items were rated on a 7-point scale.
For subjective reasonability, subjective actionability, and willingness to act, we additionally asked participants to provide an open-ended rationale for each rating.
We did not collect rationales for decision acceptance because it captures participants' overall acceptance of the decision outcome as a downstream response to the presented recourse evaluations (see SM, Section~\ref{supp_sec:correlation_benefits} for associations between decision acceptance and the other items).

For participants in multiple-option conditions, we used a selection-and-rating process for the three recourse-evaluation measures:
for each measure, participants first selected the option they considered best on that dimension, and then rated the selected option and provided its rationale.
This procedure was not applied in the single-option condition.

\subsubsection{Psychological Costs}
For cognitive load, this study used NASA-TLX~\cite{Hart1988}, which assesses subjective workload on experimental tasks with six dimensions, given its widespread use in HCI studies~\cite{Kosch2023}.
We adopted it to observe participants' cognitive load on the comprehension task.
Among the six dimensions, we used the following three:
\begin{description}
    \item[Mental demand:] the extent to which mental and perceptual activity is required for the comprehension task.
    \item[Effort:] the extent to which participants have to work hard to accomplish the comprehension task.
    \item[Frustration:] the extent to which participants feel irritated and stressed to accomplish the comprehension task.
\end{description}
These are scaled in 5-point increments from 0 to 100.
Given the context of the comprehension task with minimal physical activity, no strict time pressure, and no explicit emphasis on performance, we do not discuss the remaining dimensions---Physical Demand, Temporal Demand, and Performance (see SM, Section~\ref{supp_subsec:quantitative_full} for full results).

For emotional burdens, we observed:
\begin{description}
    \item[Negative emotional experience:] the number of emotions participants experience in reviewing the recourse options among regret, shame, guilt, self-blame, disappointment, disgust, dissatisfaction, and discrimination,
\end{description}
considering that counterfactual thinking can evoke these emotions as it highlights alternative pasts~\cite{Byrne2016}, personal shortcomings~\cite{Tominaga2024}, or insufficient achievement~\cite{Tominaga2025}.  

\subsection{Procedure\label{subsec:procedure}}
\begin{figure*}[t]
    \centering
    \includegraphics[width=1.9\columnwidth]{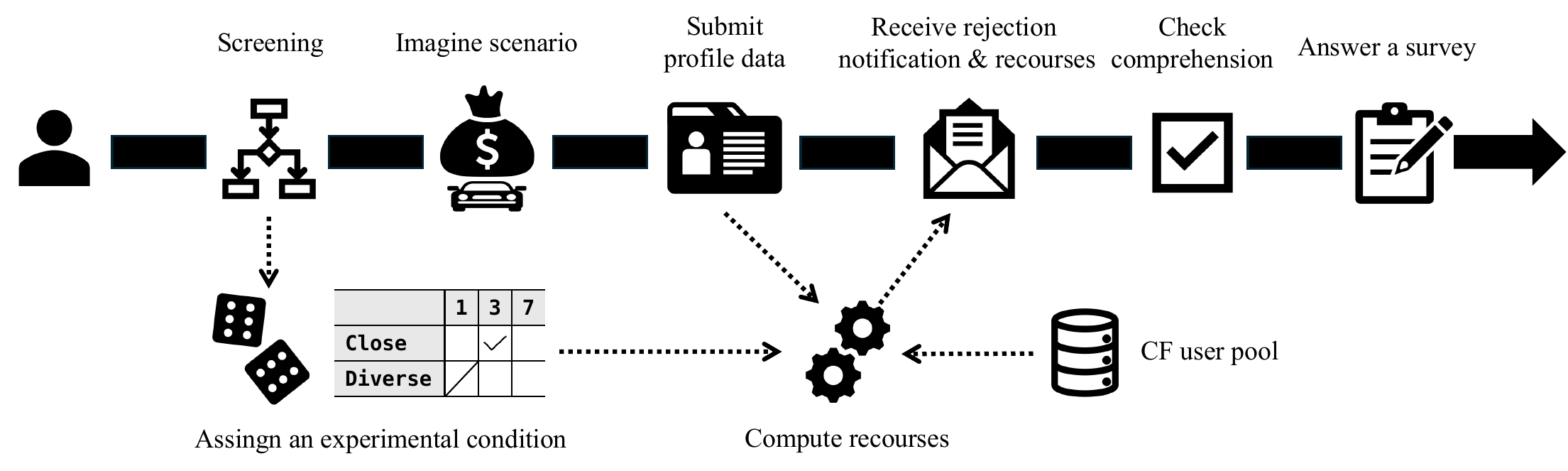}
    \caption{Overview of the experimental procedure.}
    \label{fig:procedure}
\end{figure*}

This experiment was conducted from February to March 2025 in Japan.
Figure~\ref{fig:procedure} describes the overall procedure.

\paragraph{Participants.}
Participants were recruited via the Japanese online survey company.  
They were first informed of the study's purpose and content, and consent was obtained.
To enhance ecological validity, we conducted the screening survey to select those who (1) work at a private company or public institution, (2) consider a car purchase, (3) hold no loans, and (4) earn less than 10 million JPY annually.
In total, 1062 participated in the experiment.
All were Japanese: 300 females, 762 males; 49.2 years old on average.
They received compensation equivalent to 600 JPY for the participation.

\paragraph{Recourse provision under the scenario.}
After the screening, we randomly assigned participants to one of the five experimental conditions.
They were instructed to imagine the scenario and submit their real profile data (Section~\ref{subsec:setup}).
Based on the participants' experimental condition and profile data, we computed the recourse set for each participant (Section~\ref{subsec:manipulation}).
Participants were then shown a rejection notice along with the recourse set as an AI-generated plan for future approval (see SM, Figure~\ref{supp_fig:recourse_example} for an example).

\paragraph{Comprehension check.}
We administered the comprehension task (Section~\ref{subsec:outcome_measures}); participants' responses inconsistent with the presented recourse set were judged to have insufficient understanding and were excluded from subsequent analyses. 
As a result, 750 of the 1,067 participants were retained for analysis.
Table \ref{tab:participants_by_conditions} shows the distributions of participants across the experimental conditions.
\begin{table}[t]
    \centering
    \begin{tabular}{lccc}
        \toprule
         & $k=1$ & $k=3$ & $k=7$ \\
        \midrule
        $p=\text{Close}$ & 139 & 152 & 168 \\
        $p=\text{Diverse}$ & -- & 126 & 165 \\
        \bottomrule
    \end{tabular}
    \caption{Number of participants across the experimental conditions of the recourse-set diversity $p$ and size $k$.\label{tab:participants_by_conditions}}
\end{table}

\paragraph{Outcome measurement.}
The questionnaire survey (Section~\ref{subsec:outcome_measures}) first asked about cognitive load. 
Placing it immediately after the comprehension task allowed us to more precisely capture the cognitive load required for understanding the recourse. 
Participants then completed the questionnaires on psychological benefits and emotional burdens in sequence.

\paragraph{Manipulation check.}
To confirm that our manipulations effectively altered perceptions of diversity, we asked participants with the multiple-option conditions how diverse the presented recourse set on a 7-point scale, and confirmed that our manipulation was successful using a two-way ANOVA; perceived diversity was significantly greater for Diverse than Close sets and for 7 than 3 options (diversity: $p{=}0.042$; size: $p{=}0.002$; see SM, Table~\ref{supp_tab:manipulation_check} for full statistics).

\subsection{Analysis\label{subsec:analysis}}
For each outcome of psychological benefits and costs, we conducted a series of two-way ANOVAs with recourse-set diversity and size as factors.
When the interaction was significant, we performed Tukey's HSD for post-hoc pairwise comparisons.
As diversity is undefined for a singleton set, the single-option condition was not included in the two-way ANOVAs.
Instead, we conducted Dunnett's multiple-comparison test with Close--1 as the reference, comparing it against all other conditions to quantify the effect of multiple recourse options relative to a single option.

To interpret the results qualitatively, we coded participants' open-ended responses.
One author developed the codebook iteratively, and an external collaborator independently coded a stratified random 10\% subsample per outcome.
The coders achieved substantial agreement~\cite{Landis1977}: Cohen's $\kappa\geq 0.61$.
The remaining responses were coded by the author and summarized into overarching themes.

%% file: 4_results.tex
\section{Results and Findings}
\begin{figure*}[t]
    \centering
    \includegraphics[width=1.9\columnwidth]{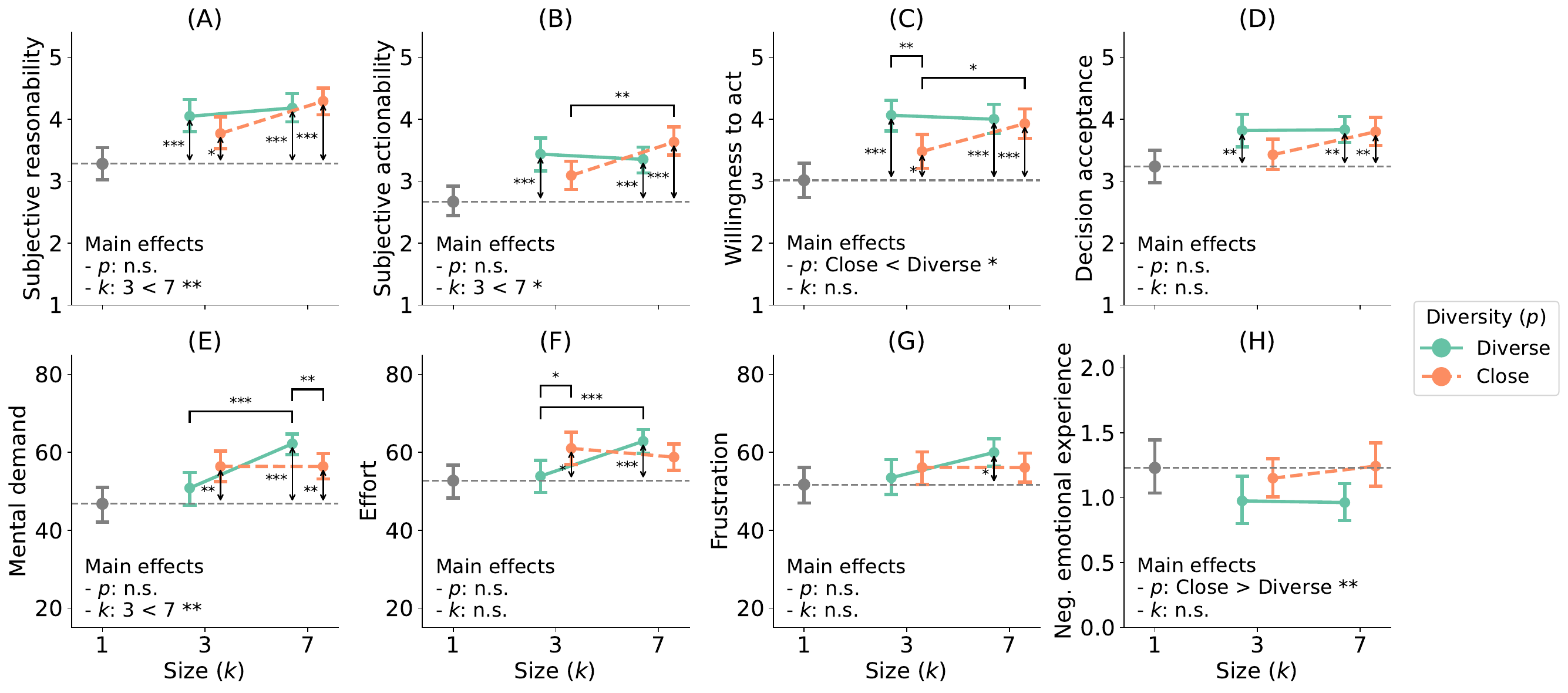}
    \caption{Psychological benefits (top row: (A)--(D)) and psychological costs (bottom row: (E)--(H)). We report (i) two-way ANOVAs with recourse-set diversity $p$ and size $k$, followed by Tukey’s HSD post-hoc tests, and (ii) Dunnett’s one-to-many comparisons using Close–1 as the reference. All $p$-values are adjusted ($^*$: $p{<}0.05$, $^{**}$: $p{<}0.01$, $^{***}$: $p{<}0.001$). Main effects from the ANOVAs are reported in the lower-left of each panel. Tukey’s HSD results are indicated by umbrella symbols, and Dunnett’s comparisons are indicated by double-arrow symbols.\label{fig:results}}
\end{figure*}

Figure~\ref{fig:results} shows how recourse-set diversity and size affect psychological benefits and costs.
Overall, while both benefits and costs tended to increase as the set size $k$ grew, the growth patterns differed between Close and Diverse.
Benefits increased up to $k{=}3$ and then plateaued in Diverse, whereas they rose more gradually with $k$ in Close and converged to Diverse at $k{=}7$ (Figure~\ref{fig:results}~(A)--(D)).
In contrast, cognitive load increased up to $k{=}3$ and then plateaued in Close, whereas they continued to grow with $k$ in Diverse, reaching a level comparable to or higher than Close at $k{=}7$ (Figure~\ref{fig:results}~(E)--(G)).
Emotional burdens were more mitigated in Diverse than Close (Figure~\ref{fig:results}~(H)).

Below, we describe three findings that unpack these patterns.
For transparency, SM reports the full quantitative outputs (Section~\ref{supp_subsec:quantitative_full}, Figure~\ref{supp_fig:tukey_all}, Tables~\ref{supp_tab:perceived_reasonability}--\ref{supp_tab:negative_emotional_experience}) and the complete qualitative results (Section~\ref{supp_subsec:qualitative_full}, Tables~\ref{supp_tab:codebook} for the codebook and \ref{supp_tab:theme_code} for the code distributions).
We refer to rationale codes of each outcome measure using specific prefixes and IDs (e.g., SR-1, SA-1, WA-1).
In this section, we focus on the key comparisons supporting Findings~1--3.

\subsection{Finding 1: Benefit of Diversification for Willingness to Act Diminishes for Larger Sets\label{subsec:finding1}}
As shown in Figure~\ref{fig:results}~(C), recourse-set diversity had a size-dependent effect on willingness to act.
A two-way ANOVA revealed a significant interaction effect of diversity and size ($F(1,607){=}4.005$, $p{=}0.046$, $\eta_p^2{=}0.007$).
Tukey's HSD post-hoc tests showed that Diverse outperformed Close only at $k{=}3$ ($p{=}0.002$), not at $k{=}7$.
Consistently with this pattern, the qualitative analysis found that Diverse--3 more frequently mentioned motives and values (WA-6; 11.1\% vs.\ 4.5\%) and less often described the recourse as unnecessary (WA-7; 0.9\% vs.\ 7.6\%) than Close--3.

Taken together, diversification can yield its benefits by helping participants to find a motivating and necessary plan at a moderate set size ($k{=}3$), whereas its advantage diminishes as the set becomes larger ($k{=}7$).

\subsection{Finding 2: Cognitive Load of Diversification Becomes Pronounced for Larger Sets\label{subsec:finding2}}
As shown in Figure~\ref{fig:results}~(E), mental demand exhibited a significant interaction ($F(1,607){=}10.121$, $p{=}0.002$, $\eta_p^2{=}0.016$).
Tukey's post-hoc comparisons indicated that Diverse exceeded Close at $k{=}7$ ($p{<}0.01$), suggesting that the cognitive cost of diversification becomes salient for larger sets.
The similar directional patterns were observed for effort and frustration: Diverse${<}$Close at $k{=}3$ and Diverse$>$Close at $k{=}7$ (Figure~\ref{fig:results}~(F), (G)).
The qualitative rationales also showed that participants' references to lack of clarity about the decision criteria were less prevalent for Diverse than Close at $k{=}3$ (SR-5; 5.4\% vs.\ 8.8\%), whereas they were slightly more prevalent for Diverse at $k{=}7$ (SR-5; 7.8\% vs.\ 6.9\%).

To understand this, we observed inconsistency across recourse options, assuming that contradictory suggestions consume cognitive resources~\cite{Kivikangas2025}.
We here defined \emph{sign conflict} as the presence of different options in a set changing the same feature in opposite directions (e.g., an option increases the number of side jobs and another decreases it in a same set).
As shown in Table~\ref{tab:descriptive_indicators}, Diverse often included sign conflict as set size increased.
\begin{table}[t]
    \setlength{\tabcolsep}{3pt}
    \centering
    \begin{tabular}{lcccc}
        \toprule
         & Close--3 & Diverse--3 & Close--7 & Diverse--7 \\
        \midrule
        Sign conflict & 16.4\% & 11.1\% & 29.2\% & 35.8\% \\
        Common change & 58.6\% & 35.7\% & 31.0\% & 27.9\% \\
        \bottomrule
    \end{tabular}
    \caption{Proportions of recourse sets with sign conflict (i.e., different options within a set change a same feature in opposite directions) or common change (i.e., all options within a set change a same feature) in each condition, as observed in exploratory analyses.}
    \label{tab:descriptive_indicators}
\end{table}
From a descriptive post-hoc analysis, we also found that adding an explanatory variable of sign conflict to the OLS model significantly improved the model fit for mental demand beyond the diversity and size (see SM, Table~\ref{supp_tab:model_comparison_indicators}).

These results and observations suggest that diversification of larger sets ($k{=}7$) often induce within-set inconsistency, possibly imposing more cognitive load.

\subsection{Finding 3: Emotional Burden Is Unlikely to Worsen under Diversification}
As shown in Figure~\ref{fig:results}~(H), negative emotional experience was reduced under diversification regardless of set size ($F(1,607){=}7.949$, $p{=}0.005$, $\eta_p^2{=}0.013$).
This pattern runs counter to a naive expectation that increasing the number or diversity of options would amplify emotional burden (Section~\ref{sec:related_work}), and it suggests that emotional burdens behave differently from cognitive costs (Finding~2).

To interpret this, we focused on redundancy among recourse options, inspired by the idea that receiving the same advice repeatedly is unpleasant~\cite{Cacioppo1979,Kocielnik2017}.
As its proxy, we explored \emph{common change}, defined as the presence of a feature that is changed in a same direction by all options within a set.
As shown in Table~\ref{tab:descriptive_indicators}, Diverse tended to exhibit fewer common changes than Close, especially at $k{=}3$.
However, a descriptive post-hoc analysis showed that common change did not relate to negative emotional experience (see SM, Table~\ref{supp_tab:model_comparison_indicators}).

To sum up, diversification was unlikely to exacerbate emotional burden; the drivers of this pattern warrant further investigation, including redundancy.

%% file: 5_discussion.tex
\section{Discussion}
As indicated by our findings, diversification can increase willingness to act for smaller sets, while cognitive load likely become salient for larger sets.
This size-dependent pattern warns failure modes in generating large diverse sets and motivates design strategies to address them.
This section describes its details, and then outlines limitations and future work.

\subsection{Implications}
The size-dependent trade-off between benefits and costs rests on two principles of diversification.
First, as set size grows, the relative advantage of diversification shrinks.
This is because, even when different policies are used (e.g., Close or Diverse), each chooses increasingly overlapping counterfactual samples.
This convergence is likely as promising counterfactuals are scarce in practice~\cite{Keane2020}.
This principle is consistent with our result that the diversification benefits diminished for larger sets (Section~\ref{subsec:finding1}).

Second, diversification has a structural property that prioritizes counterfactuals altering different features, and once these are exhausted, then chooses samples altering same features in opposite directions.
This is optimal strategy to minimize the objective function in Equation~(\ref{eq:computing_recourse_set}).
Consequently, inconsistent suggestions (e.g., sign conflict) likely emerge in large diverse recourse sets.
Given that decision making under conflicting aims increases cognitive effort~\cite{Kivikangas2025}, sign conflict possibly contributed to the increased mental demand (Section~\ref{subsec:finding2}).
Notably, if the first principle was only considered, differences in cognitive load between the diversity conditions would shrink as set size grows.
However, we observed that, rather than converging, cognitive load became more salient under diversification for larger sets.

These interpretations suggest that diversification may raise cognitive load when set size is ignored, calling for remedies to both the scarcity of promising samples and inconsistencies across recourse options.
This perspective lends empirical support to prior technical directions, including mechanisms for securing sufficiently many promising samples in model construction~\cite{Kanamori2024} or securing coherency among recourse options~\cite{Rasouli2024}.

Close sets increased benefits with set size, but plateaued cognitive load (Figure~\ref{fig:results}), suggesting that multiple, similar but distinct suggestions may increase perceived persuasiveness.
Since sets were constructed from unique profiles (Section~\ref{subsubsec:recourse_computation}), no sets presented duplicate options.
Instead, repeating identical recourse options would likely attenuate the effect due to repetition discomfort~\cite{Cacioppo1979}.

\subsection{Recommendations\label{subsec:recommendation}}
Our findings inform that the policy of recourse diversification algorithms in AI systems should consider the recourse-set size and properties.
For relatively small sets, diversification is recommended because it can increase willingness to act while keeping cognitive and emotional burdens manageable.
For relatively large sets, by contrast, it is desirable to suppress within-set inconsistency and determine the set size adaptively.
Following this policy, we recommend extending the proximity–diversity objective by introducing set size and within-set inconsistency as either hard constraints or penalty terms, where inconsistency is captured by sign conflicts across recourse options.

\subsection{Limitations and Future Directions}
Our study adopted set size $k{=}1,3,7$ based on cognitive theory to create qualitatively distinct conditions as noted in Section~\ref{subsubsec:factor}, but we did not test intermediate sizes such as $k{=}5$; thus, we cannot precisely localize the set size at which the benefit-cost balance turns. 
Importantly, our contribution does not lie in the specific numbers per se (e.g., 3 or 7), but in elucidating the qualitative pattern that the benefits of diversification can plateau beyond a certain size while the costs continue to grow with $k$. 
Accordingly, $k{=}1,3,7$ should not be interpreted as generalizable ``magic numbers,'' as the turning point may vary with the domain and dataset, and ultimately with the availability of promising counterfactual samples.
To identify this turning point in practice, future work should calibrate it within the target task.
Since our results suggest that costs increase with set size, such efforts should prioritize inspecting the plateau point of benefits.

We used an automobile-loan setting as the experimental task; therefore, there remains room for future work to examine whether our findings can be applied to different decision-making domains.
In addition, our sample was skewed toward men and drawn from a single cultural context.
As decision-making styles~\cite{Hofstede2011} and trust in AI~\cite{Maslej2025} may vary across cultures, further studies should test whether similar results hold among decision subjects with greater demographic and cultural diversity.

Translating our empirical findings into concrete technical solutions remains future work.
As noted in Section~\ref{subsec:recommendation}, one promising direction is to incorporate set properties (e.g., size or sign conflict) into recourse generation objective.
These directions would help establish generation and presentation techniques that automatically tune the benefit-cost balance.

%% file: 6_conclusion.tex
\section{Conclusion}
Through a randomized controlled between-subjects experiment ($N{=}750$) complemented with quantitative and qualitative analyses, this paper provides empirical evidence on the size-dependent trade-off between psychological benefits and costs of diversifying algorithmic recourse.
Specifically, diversification increased willingness to act for smaller sets, whereas it made cognitive load more pronounced for larger sets.
Our post-hoc analysis suggests a key pitfall: diversifying larger sets often induces inconsistencies across options, which may amplify cognitive load.
To address this, we highlighted the potential of diagnostic properties for recourse verification and optimization. 
We hope these foundational findings and discussions will pave the way for human-centered AI systems integrating human cognition and psychology.

%% file: 7_supplementary.tex
\appendix
\clearpage      
\onecolumn      

\setcounter{figure}{0} 
\setcounter{table}{0}

\renewcommand{\thefigure}{S\arabic{figure}}
\renewcommand{\thetable}{S\arabic{table}}

\newlength{\cellw}
\setlength{\cellw}{0.25\textwidth}

\newlength{\cellh}
\setlength{\cellh}{1.5\cellw}

\newcolumntype{C}[1]{>{\centering\arraybackslash}p{#1}}


\begin{center}
{\Large\bfseries Supplementary Materials}\par
\end{center}

\section{Greedy Algorithm for Computing Recourse Sets\label{supp_sec:alg}}
\begin{algorithm}[h]
    \caption{Greedy solver for the recourse-set optimization}
    \label{alg:greedy-recourse}
    \begin{algorithmic}[1]
        \Require Candidate set $\mathit{Candidates}$, participant $\bm{x}$, target size $k$, criterion $p \in \{\text{Close}, \text{Diverse}\}$
        \Ensure Selected counterfactual set $C$ of size $k$
        \State $C \gets \emptyset$
        \State $\mathit{Available} \gets \mathit{Candidates} \cap \mathcal{A} \cap \mathcal{Z}$ \Comment{Feasible search region}
        \Statex
        \Comment{Initialization: pick the closest sample to $\bm{x}$}
        \State $\bm{c}^\star \gets \arg\min_{\bm{c}\in \mathit{Available}} \ \text{dist}(\bm{c}, \bm{x})$
        \State $C \gets C \cup \{\bm{c}^\star\}$,\quad $\mathit{Available} \gets \mathit{Available} \setminus \{\bm{c}^\star\}$
        \Statex
        \Comment{Iteratively add samples that greedily minimize the objective}
        \While{$|C| < k$}
            \State $\bm{c}^\dagger \gets \arg\min_{\bm{c}\in \mathit{Available}} \ J\!\left(C \cup \{\bm{c}\}; \bm{x}, p\right)$
            \State $C \gets C \cup \{\bm{c}^\dagger\}$,\quad $\mathit{Available} \gets \mathit{Available} \setminus \{\bm{c}^\dagger\}$
        \EndWhile
        \State \textbf{return} $S$
        \Statex
        \Function{$J$}{$C$; $\bm{x}$, $p$} \Comment{Objective on current set $C$}
            \State $m \gets |C|$ \Comment{Current set size}
            \State $\text{avg\_px} \gets \frac{1}{m} \sum_{\bm{c}\in C} \text{dist}(\bm{c}, \bm{x})$
            \If{$p=\text{Close}$}
                \State \textbf{return} $\text{avg\_px}$
            \Else \Comment{$p=\text{Diverse}$}
                \If{$m=1$}
                    \State \textbf{return} $\text{avg\_px}$ \Comment{No pair to measure dissimilarity}
                \Else
                    \State $\text{avg\_pp} \gets \frac{2}{m(m-1)} \sum_{\substack{\bm{c}^{(i)},\,\bm{c}^{(j)} \in C \\ i<j}} \text{dist}\!\left(\bm{c}^{(i)}, \bm{c}^{(j)}\right)$
                    \State \textbf{return} $\text{avg\_px} - \text{avg\_pp}$
                \EndIf
            \EndIf
        \EndFunction
    \end{algorithmic}
\end{algorithm}

\clearpage
\section{Questionnaire Surveys}\label{supp_sec:survey}
\subsection{Screening Survey}\label{supp_subsec:screening}
\begin{enumerate}
    \item Please tell us your age.
    \\ (\ \ \ \ \ ) years old
    \item Please tell us your gender.
    \\ 1. Female
    \\ 2. Male
    \\ 3. Other
    \item Please tell us your occupation.
    \\ 1. Company worker (full-time)
    \\ 2. Company worker (contract)
    \\ 3. Company worker (temporary)
    \\ 4. Company worker (part-time)
    \\ 5. Government worker
    \\ 6. Self-employed / Freelance
    \\ 7. Homemaker
    \\ 8. Part-time job
    \\ 9. Student
    \\ 10. Unemployed / Retired
    \\ 11. Other
    \item Please tell us your annual income.
    \\ 1. - 2M
    \\ 2. 2M - 3M
    \\ 3. 3M - 4M
    \\ 4. 4M - 5M
    \\ 5. 5M - 6M
    \\ 6. 6M - 7M
    \\ 7. 7M - 8M
    \\ 8. 8M - 9M
    \\ 9. 9M - 10M
    \\ 10. 10M -
    \item Please tell us about your thoughts regarding the purchase of a car.
    \\ 1. I plan to purchase within a year
    \\ 2. I plan to purchase within three years
    \\ 3. I plan to purchase within five years
    \\ 4. While the timing is undecided, I plan to purchase
    \\ 5. I do not plan to purchase
    \item Please select all of the following that apply to your current loan situation.
    \\ 1. Card loans
    \\ 2. Automibile loans
    \\ 3. Mortage loans
    \\ 4. Education loans
    \\ 5. Free loans
    \\ 6. Business loans
    \\ 7. Other loans
    \\ 8. I do not have any loans
\end{enumerate}

\clearpage
\subsection{Profile Data Survey}\label{supp_subsec:profile_data_survey}
Please respond to the following questions, imaging the scenario below.

\noindent
Currently, you are considering borrowing a two-year car loan to purchase a car equivalent to one-third of your annual income.
\begin{figure}[h]
    \centering
    \includegraphics[width=0.65\linewidth]{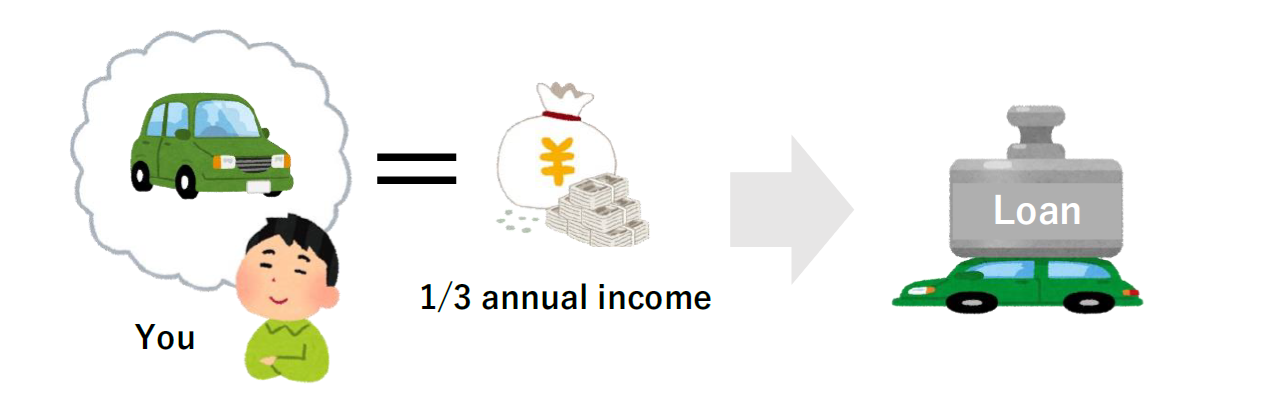}
    \label{fig:profile_1}
\end{figure}

\noindent
Right now, you are at a financial institution undergoing a loan assessment. In this assessment, an AI is determining the approval or denial of the loan based on the applicant's profile data.
\begin{figure}[h]
    \centering
    \includegraphics[width=0.65\linewidth]{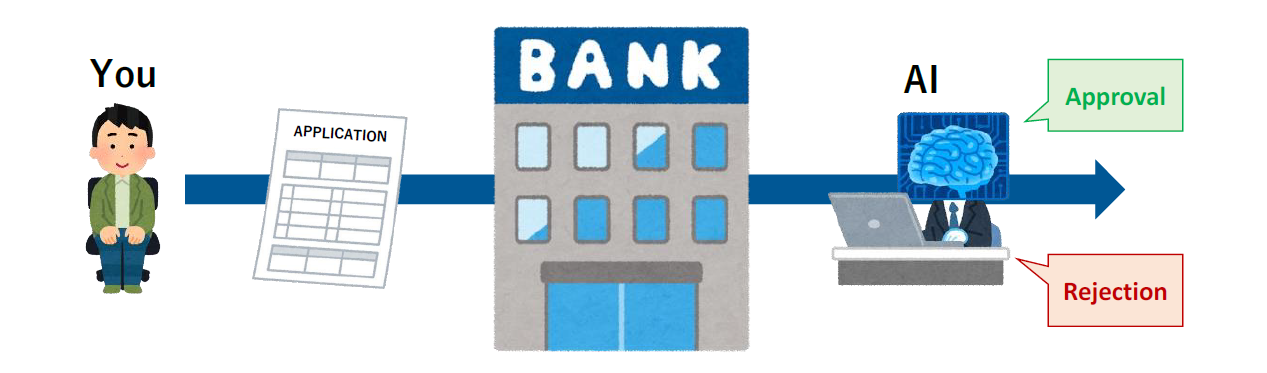}
    \label{fig:profile_2}
\end{figure}

\noindent
To undergo this evaluation, you have decided to submit your own profile data.
\begin{figure}[h]
    \centering
    \includegraphics[width=0.65\linewidth]{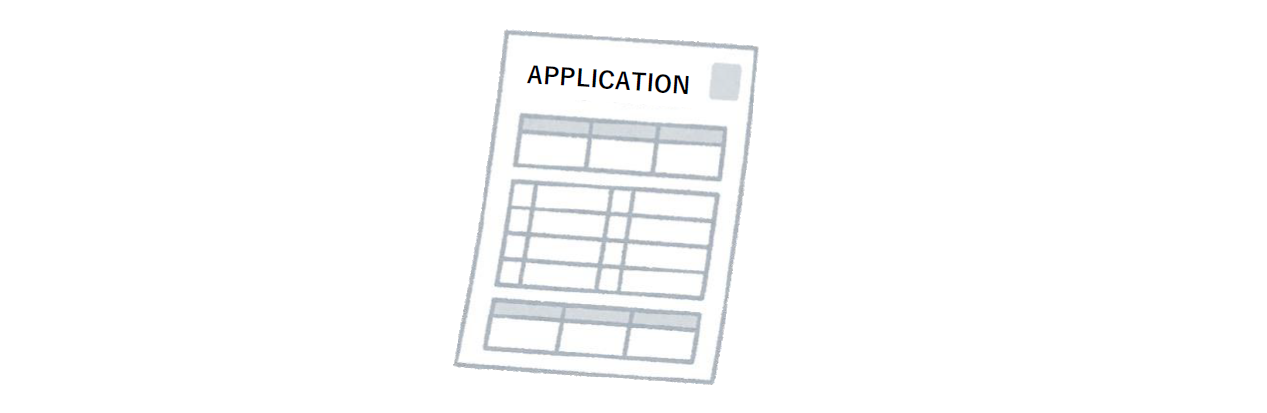}
    \label{fig:profile_3}
\end{figure}

\clearpage
\noindent
Please select the option that applies to you for each of the following items.
\begin{table}[h]
  \centering
  \begin{tabular}{clp{0.65\linewidth}}
    \toprule
    \# & Profile item (feature) & Option\\
    \midrule
    1 & Residence & 1. Tokyo / 2. Other than Tokyo\\
    2 & Type of residence & 1. Owned house / 2. Rental housing\\
    3 & Education & 1. High school / 2. Junior college / 3. University (bachelor) / 4. Graduate school (master) / 5. Graduate school (doctor) \\
    4 & Employment & 1. Private company / 2. Public institution \\
    5 & Job title & 1. Employee / 2. Supervisor / 3. Section head / 4. Section chief / 5. Assistant general manager / 6. Manager / 7. General manager / 8. Executive director / 9. Senior executive director / 10. President \\
    6 & Years of service & 1. 0-1 year / 2. 1-3 years / 3. 3-5 years / 4. 5-10 years / 5. 10-20 years / 6. 20- years \\
    7 & Management career & 1. No / 2. 0-1 year / 3. 1-3 years / 4. 3-5 years / 5. 5-10 years / 6. 10-20 years / 7. 20- years \\
    8 & Daily working hours & 1. 0-2 hours / 2. 2-4 hours / 3. 4-6 hours / 4. 6-8 hours / 5. 8-10 hours / 6. 10-12 hours / 7. 12- hours \\
    9 & Daily remote working hours & 1. 0-2 hours / 2. 2-4 hours / 3. 4-6 hours / 4. 6-8 hours / 5. 8-10 hours / 6. 10-12 hours / 7. 12- hours \\
    10 & Number of side jobs & 1. No / 2. 1 job / 3. 2 jobs / 4. 3 jobs / 5. 4 jobs / 6. 5- jobs \\
    11 & Job change experience & 1. No / 2. Yes \\
    12 & Overseas work experience & 1. No / 2. Yes \\
    13 & Study abroad experience & 1. No / 2. Yes \\
    14 & Best TOEIC$^\dagger$ score & 1. No / 2. 10-400 / 3. 400-495 / 4. 500-595 / 5. 600-695 / 6. 700-795 / 7. 800-895 / 8. 900-990 \\
    15 & Facebook use & 1. No / 2. Yes \\
    16 & LinkedIn use & 1. No / 2. Yes \\
    \bottomrule
  \end{tabular}
   \caption{Answer options for each profile item}\label{supp_tab:profile_data}
\end{table}

\subsection{Rejection Notification and Recourse Provision}\label{supp_subsec:notification_provision}
You had come for a two-year loan assessment with the intention of purchasing a car equivalent to one-third of your annual income. 
But unfortunately, you were declined in the AI-based loan assessment.
\begin{figure}[H]
    \centering
    \includegraphics[width=0.65\linewidth]{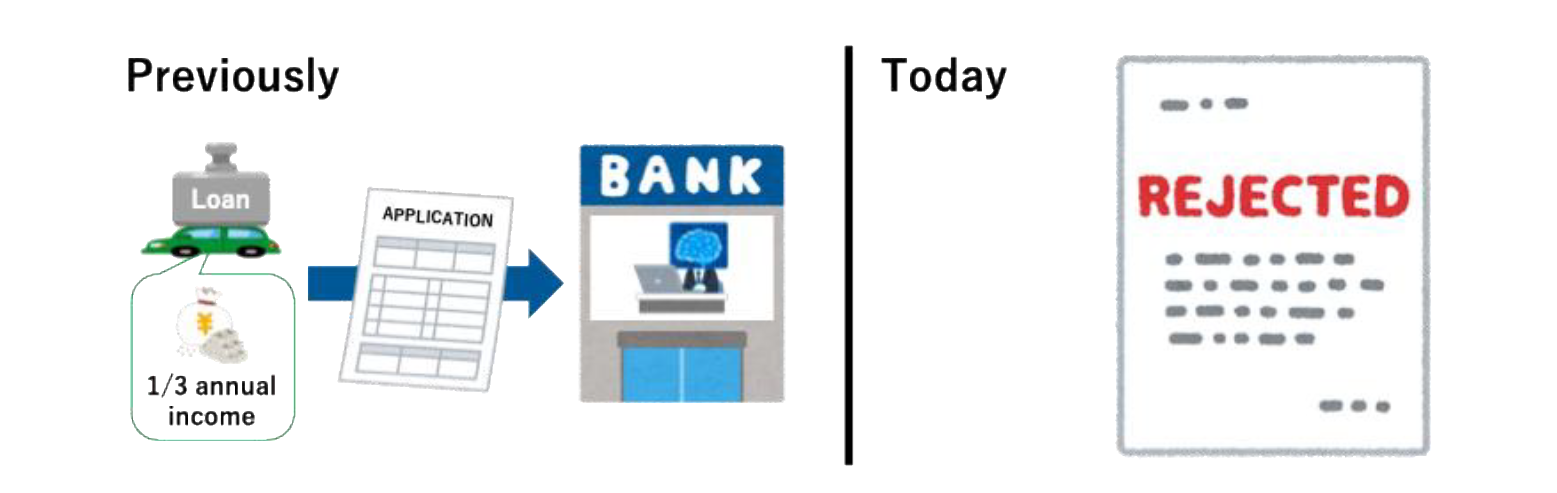}
    \label{fig:notification_1}
\end{figure}

\noindent
The AI system used in this assessment can provide a failed applicant with an ideal profile for her/his current profile, like the figure below, as a plan of actions to ensure that her/his next application will be approved.
\begin{figure}[H]
    \centering
    \includegraphics[width=0.65\linewidth]{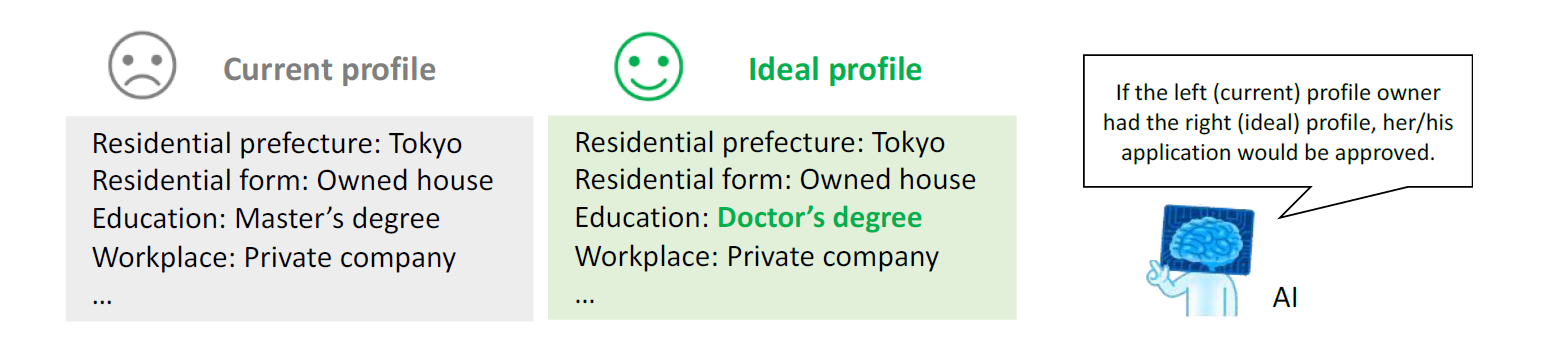}
    \label{fig:notification_2}
\end{figure}

\noindent
If you understand it, please proceed to the next page.

\noindent
The AI provides several options for action plans for you, as shown below.
Where possible, it suggests plans with lower change costs in order to reduce your burden\footnote{During all the measurements (from Section~\ref{supp_subsec:comprehension_check} to \ref{supp_subsec:measure_negative_emotional_experience}), the recourse set was presented to participants.}.
\begin{figure}[ht]
  \centering
  \begin{subfigure}[t]{\cellw}
    \centering
    \begin{minipage}[c][\cellh][c]{\linewidth}
      \centering
      \includegraphics[width=\linewidth, max height=\cellh, keepaspectratio, valign=m]{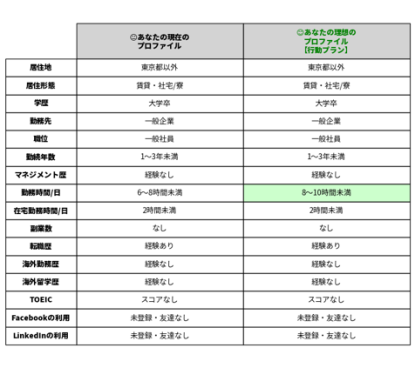}
    \end{minipage}
    \caption{Close--1}
  \end{subfigure}\hspace{5mm}
  \begin{subfigure}[t]{\cellw}
    \centering
    \begin{minipage}[c][\cellh][c]{\linewidth}
      \centering
      \includegraphics[width=\linewidth, max height=\cellh, keepaspectratio, valign=m]{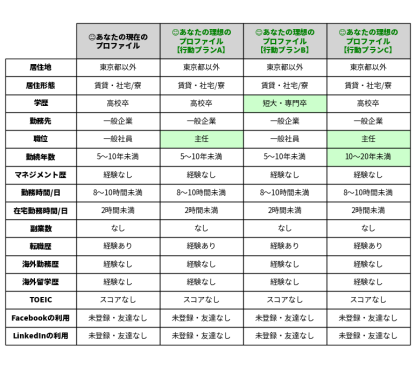}
    \end{minipage}
    \caption{Close--3}
  \end{subfigure}\hspace{5mm}
  \begin{subfigure}[t]{\cellw}
    \centering
    \begin{minipage}[c][\cellh][c]{\linewidth}
      \centering
      \includegraphics[width=\linewidth, max height=\cellh, keepaspectratio, valign=m]{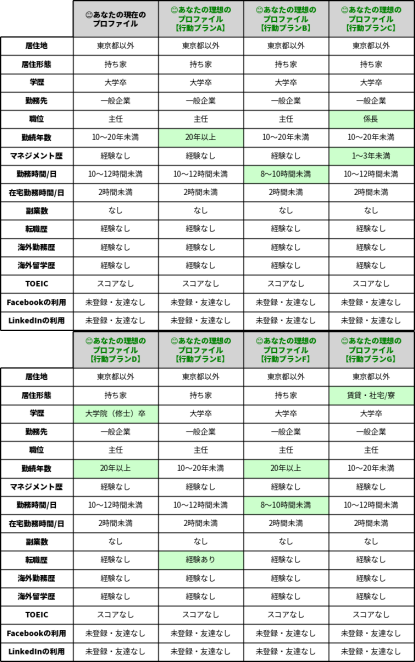}
    \end{minipage}
    \caption{Close--7}
  \end{subfigure}

  \vspace*{5mm}

  \hspace{\cellw}\hspace{5mm}
  \begin{subfigure}[t]{\cellw}
    \centering
    \begin{minipage}[c][\cellh][c]{\linewidth}
      \centering
      \includegraphics[width=\linewidth, max height=\cellh, keepaspectratio, valign=m]{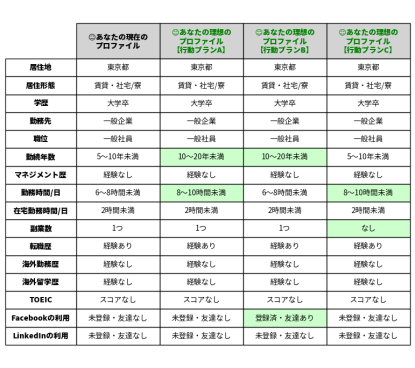}
    \end{minipage}
    \caption{Diverse--3}
  \end{subfigure}\hspace{5mm}
  \begin{subfigure}[t]{\cellw}
    \centering
    \begin{minipage}[c][\cellh][c]{\linewidth}
      \centering
      \includegraphics[width=\linewidth, max height=\cellh, keepaspectratio, valign=m]{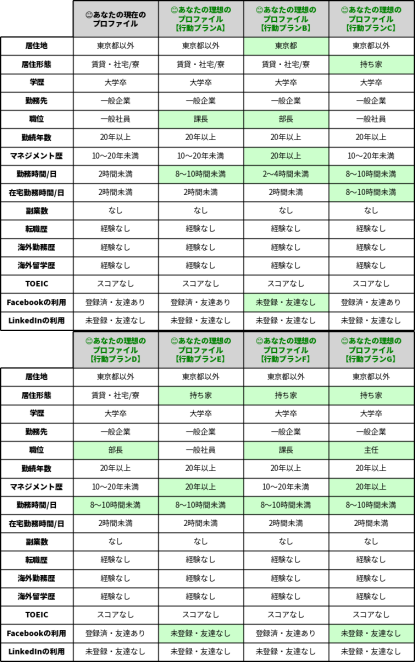}
    \end{minipage}
    \caption{Diverse--7}
  \end{subfigure}
  
  \caption{Examples of the recourse sets shown to participants under each experimental condition (one example per condition). Each subfigure shows the feature names in the leftmost column (i.e., index), shows the focal participant’s profile (top-left for Close–7 and Diverse–7) in the next column (i.e., the leftmost column within the table body), and the profiles of the counterfactual samples in the remaining columns. Entries that differ from the participant's profile are highlighted in green. Counterfactual samples were selected so that no two samples within the same set are identical. The study used a between-subjects design: each participant was assigned to exactly one of the five conditions and evaluated only the corresponding recourse set. This figure therefore provides one illustrative example per condition, and the specific set shown varies across participants.}\label{supp_fig:recourse_example}
\end{figure}

\subsection{Comprehension Check}\label{supp_subsec:comprehension_check}
The AI systems generated the plan(s) to achieve the following two objectives:
\begin{enumerate}
    \item that you understand the reason why the AI systems made the negative decision.
    \item that you understand the necessary actions to obtain the positive decision.
\end{enumerate}
Please complete the following tasks in order to extract these information from the plan(s) presented.
\\

\begin{description}
    \item[Task 1a] Based on the above plan(s), please state one reason why your application was rejected by the AI.
    \\ ({\color{gray} e.g., Because the type of residence is ``rental housing,'' not ``owned home''}\hfill)
    \item[Task 1b] Among other reasons for the rejection, please select up to three you believe had the strongest influence.
    \\ $\square$ Residence
    \\ $\square$ Type of residence
    \\ $\square$ Education
    \\ $\square$ Employment
    \\ $\square$ Job title
    \\ $\square$ Years of service
    \\ $\square$ Management career
    \\ $\square$ Daily working hours
    \\ $\square$ Daily remote working hours
    \\ $\square$ Number of side jobs
    \\ $\square$ Job change experience
    \\ $\square$ Overseas work experience
    \\ $\square$ Study abroad experience
    \\ $\square$ Best TOEIC score
    \\ $\square$ Facebook use
    \\ $\square$ LinkedIn use
    \\ $\square$ None of the above
\end{description}
\vspace{2mm}
\begin{description}
    \item[Task 2a] Based on the above plan(s), please state one action you need to take for your application to be approved.
    \\ ({\color{gray} e.g., Change the type of residence from ``rental housing'' to ``owned home''}\hfill)
    \item[Task 2b] Among other recommended actions, please select up to three you believe would be most effective.
    \\ $\square$ Residence
    \\ $\square$ Type of residence
    \\ $\square$ Education
    \\ $\square$ Employment
    \\ $\square$ Job title
    \\ $\square$ Years of service
    \\ $\square$ Management career
    \\ $\square$ Daily working hours
    \\ $\square$ Daily remote working hours
    \\ $\square$ Number of side jobs
    \\ $\square$ Job change experience
    \\ $\square$ Overseas work experience
    \\ $\square$ Study abroad experience
    \\ $\square$ Best TOEIC score
    \\ $\square$ Facebook use
    \\ $\square$ LinkedIn use
    \\ $\square$ None of the above
\end{description}

\subsection{Measurement of Cognitive Load}\label{supp_subsec:measure_cognitive_load}
Next, we would like to ask about your perceived workload while completing the task you just performed\footnote{We used a Japanese version of the NASA-TLX adapted for this experiment. The items were originally presented in Japanese, but for convenience we provide their English translations below. Participants answered each item on a 21-point scale ranging from 0 (left anchor) to 100 (right anchor) in increments of 5.}.
\begin{itemize}
    \item \textbf{Mental demand}.
    \\ When performing these tasks, how much mental and perceptual activity  was required? (i.e.., cognitive processing for decision making or information processing such as thinking, deciding, calculating, remembering, etc.)
    \\ 0. Very low -- 100. Very high
    \item \textbf{Physical demand}.
    \\ When performing these tasks, how much physical activity was required? (i.e., physical actions or operations such as clicking, scrolling, zooming, standing, sitting, etc.)
    \\ 0. Very low -- 100. Very high
    \item \textbf{Temporal demand}.
    \\ When performing these tasks, how much time pressure did you feel?
    \\ 0. Very low -- 100. Very high
    \item \textbf{Performance}.
    \\ When performing these tasks, how successful do you think you were in accomplishing the goals of the task?
    \\ 0. Failure -- 100. Perfect\footnote{In the original NASA-TLX, the Performance scale is anchored such that 0 corresponds to Perfect and 100 corresponds to Failure. To maintain consistency with the other items and to simplify interpretation, we reversed this scale. In our study, 0 indicated Failure and 100 indicated Perfect. Scores were treated accordingly in all analyses.}.
    \item \textbf{Effort}. 
    \\ When performing these tasks, how hard did you have to work to accomplish these tasks? (i.e., mental and physical effort needed)
    \\ 0. Very low -- 100. Very high
    \item \textbf{Frustration}.
    \\ When performing these tasks, how irritated or stressed did you feel?
    \\ 0. Very low -- 100. Very high
\end{itemize}

\subsection{Measurement of Psychological Benefits}\label{supp_subsec:measure_psychological_benefits}
Here, we would like to ask about your impressions of the action plans suggested by the AI.

\subsubsection{For Participants Assigned to Single-option Condition (i.e., Close--1)}
\begin{itemize}
    \item \textbf{Perceived reasonability}.
    \begin{itemize}
        \item To what extent do you think the action plan is reasonable as an explanation for the rejection?
        \\ 1. Strongly disagree -- 7. Strongly agree
        \item Why do you think so?
        \\ (\hfill)
    \end{itemize}
    \item \textbf{Perceived actionability}.
    \begin{itemize}
        \item To what extent is it difficult or easy for you to carry out the action plan?
        \\ 1. Very difficult -- 7. Very easy
        \item Why do you think so?
        \\ (\hfill)
    \end{itemize}
    \item \textbf{Willingness to act}.
    \begin{itemize}
        \item To what extent are you willing to carry out the action plan?
        \\ 1. Not at all willing -- 7. Very willing
        \item Why do you think so?
        \\ (\hfill)
    \end{itemize}
    \item \textbf{Decision acceptance}.
    \begin{itemize}
        \item After reviewing the action plan suggested by the AI, to what extent do you accept the decision outcome of your loan application?
        \\ 1. Cannot accept at all -- 7. Fully accept
    \end{itemize}
\end{itemize}

\subsubsection{For Participants Assigned to Multiple-option Conditions (i.e., Close--3, Diverse--3, Close--7, Diverse--7)}

\begin{itemize}
    \item \textbf{Subjective reasonability}.
    \begin{itemize}
        \item Please select the action plan that you think is the most reasonable explanation for why your loan application was rejected.
        \\ For Close--3, Diverse--3: Plan A, Plan B, Plan C
        \\ For Close--7, Diverse--7: Plan A, Plan B, Plan C, Plan D, Plan E, Plan F, Plan G
        \item To what extent do you think the action plan you selected is reasonable as an explanation for the rejection?
        \\ 1. Strongly disagree -- 7. Strongly agree
        \item Why do you think so?
        \\ (\hfill)
    \end{itemize}
    \item \textbf{Subjective actionability}.
    \begin{itemize}
        \item Please select the action plan that would be the easiest for you to carry out.
        \\ For Close--3, Diverse--3: Plan A, Plan B, Plan C
        \\ For Close--7, Diverse--7: Plan A, Plan B, Plan C, Plan D, Plan E, Plan F, Plan G
        \item To what extent is it difficult or easy for you to carry out the action plan?
        \\ 1. Very difficult -- 7. Very easy
        \item Why do you think so?
        \\ (\hfill)
    \end{itemize}
    \item \textbf{Willingness to act}.
    \begin{itemize}
        \item Please select the action plan that you would most like to carry out.
        \\ For Close--3, Diverse--3: Plan A, Plan B, Plan C
        \\ For Close--7, Diverse--7: Plan A, Plan B, Plan C, Plan D, Plan E, Plan F, Plan G
        \item To what extent are you willing to carry out the action plan?
        \\ 1. Not at all willing -- 7. Very willing
        \item Why do you think so?
        \\ (\hfill)
    \end{itemize}
    \item \textbf{Decision acceptance}.
    \begin{itemize}
        \item After reviewing the action plans suggested by the AI, to what extent do you accept the decision outcome of your loan application?
        \\ 1. Cannot accept at all -- 7. Fully accept
    \end{itemize}
\end{itemize}

\subsection{Measurement of Negative Emotional Experience}\label{supp_subsec:measure_negative_emotional_experience}
Next, we would like to ask about your emotional experience related to the action plans suggested by the AI.

\begin{itemize}
    \item \textbf{Negative emotional experience.}
    \begin{itemize}
        \item How did you feel when reviewing the action plans suggested by the AI? Please select all that apply from the options below.
        \\ $\square$ Regret (e.g., thought that you should or should not have done something in the past)
        \\ $\square$ Shame (e.g., felt embarrassed because shortcomings were pointed out)
        \\ $\square$ Guilt (e.g., felt that your actions might cause trouble for others)
        \\ $\square$ Self-blame (e.g., felt like blaming your past self)
        \\ $\square$ Disappointment (e.g., felt let down because it was not as expected)
        \\ $\square$ Disgust (e.g., felt a negative impression toward the evaluation or AI)
        \\ $\square$ Dissatisfaction (e.g., felt dissatisfied because features you did not want to be used, or features unfavorable to you, were applied)
        \\ $\square$ Discrimination (e.g., felt that the content was discriminatory)
        \\ $\square$ None of the above
    \end{itemize}
\end{itemize}

\subsection{Measurement of Perceived Diversity for Manipulation Check}\label{supp_subsec:measure_perceived_diversity}
\begin{itemize}
    \item \textbf{Perceived diversity.}
    \begin{itemize}
        \item Do you think the proposed action plans, as a whole, explain the decision outcome from diverse perspectives?
        \\ 1. Strongly disagree -- 7. Strongly agree
    \end{itemize}
\end{itemize}

\clearpage
\section{Correlations Among Measured Outcomes of Psychological Benefits\label{supp_sec:correlation_benefits}}
\begin{figure}[h]
    \centering
    \begin{subfigure}{0.33\linewidth}
        \centering
        \includegraphics[width=\linewidth]{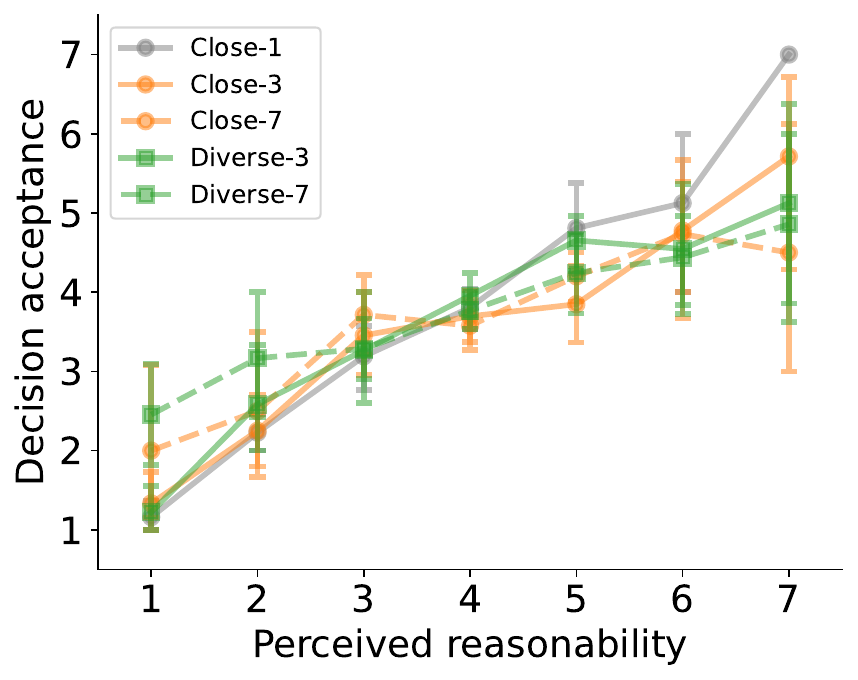}
        \caption{Perceived reasonability}
    \end{subfigure}
    \hfill
    \begin{subfigure}{0.33\linewidth}
        \centering
        \includegraphics[width=\linewidth]{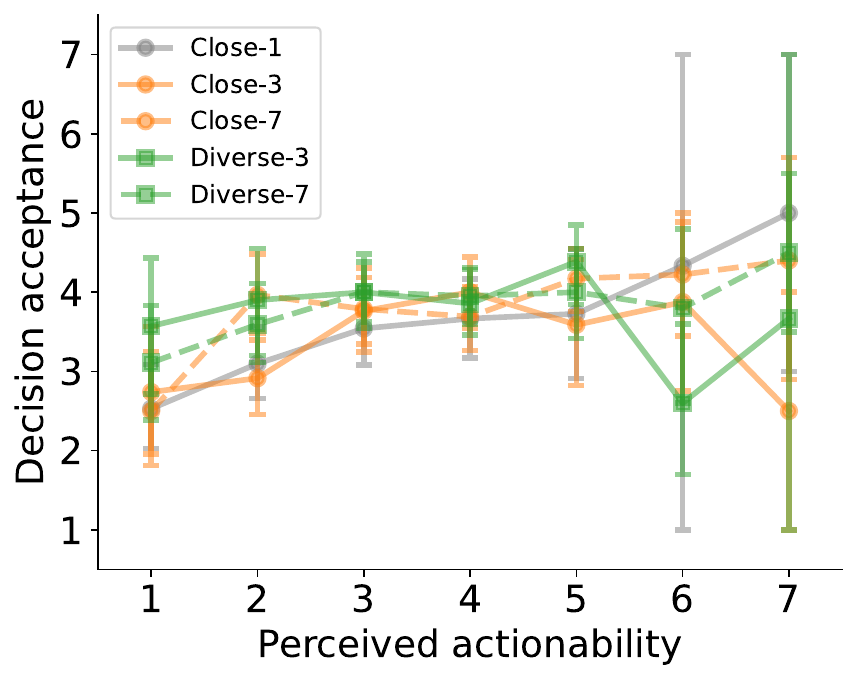}
        \caption{Perceived actionability}
    \end{subfigure}
    \hfill
    \begin{subfigure}{0.33\linewidth}
        \centering
        \includegraphics[width=\linewidth]{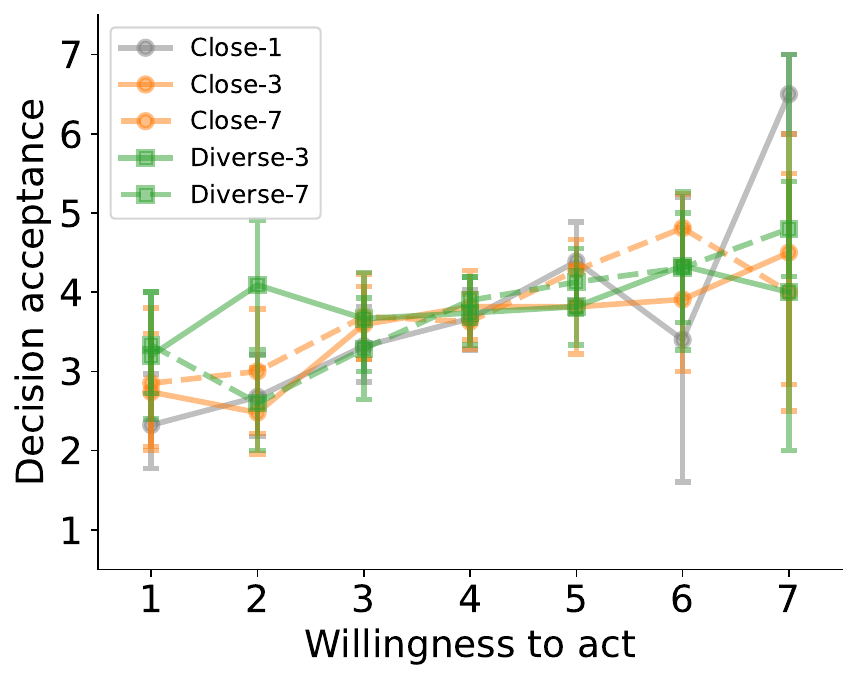}
        \caption{Willingness to act}
    \end{subfigure}
    \caption{Relationships between decision acceptance and the other measured outcomes of psychological benefits}
    \label{supp_fig:correlation_benefits}
\end{figure}

We examined the relationships between decision acceptance and the other psychological benefits. 
Figures~\ref{supp_fig:correlation_benefits} visualize how decision acceptance varies with (a) perceived reasonability, (b) perceived actionability, and (c) willingness to act.
To quantify these relationships, we computed the correlations between decision acceptance and each of these measures.
Positive correlations were observed in all conditions except for the Diverse--3 condition for perceived actionability and willigness to act. Detailed statistics are reported in Table~\ref{supp_tab:correlation_benefits}.
The result that perceived reasonability is more strongly correlated with decision acceptance than perceived actionability is consistent with prior work~\cite{Tominaga2025}.
\begin{table}[h]
    \centering
    \begin{tabular}{lrrrrrr}
        \toprule
         & \multicolumn{2}{c}{Perceived reasonability} & \multicolumn{2}{c}{Perceived actionability} & \multicolumn{2}{c}{Willingness to act}\\
        \cmidrule(lr){2-3}
        \cmidrule(lr){4-5}
        \cmidrule(lr){6-7}
        Condition & Pearson's $r$ & $p$ value & Pearson's $r$ & $p$ value & Pearson's $r$ & $p$ value \\
        \midrule
        Close--1 & 0.824 & $<$.001 & 0.313 & $<$.001 & 0.461 & $<$.001\\
        Close--3 & 0.610 & $<$.001 & 0.212 & 0.009 & 0.327 & $<$.001\\
        Close--7 & 0.432 & $<$.001 & 0.219 & 0.004 & 0.348 & $<$.001\\
        Diverse--3 & 0.623 & $<$.001 & $-$0.061 & 0.500 & 0.115 & 0.200\\
        Diverse--7 & 0.428 & $<$.001 & 0.180 & 0.021 & 0.307 & $<$.001\\
        \bottomrule
    \end{tabular}
    \caption{Statistics obtained from Pearson's correlation analysis: decision acceptance vs. perceived reasonability, perceived actionability, and willingness to act across conditions.}
    \label{supp_tab:correlation_benefits}
\end{table}

\section{Quantitative Analysis and Results}\label{supp_subsec:quantitative_full}
\begin{figure}[t]
    \centering
    \includegraphics[width=0.9\linewidth]{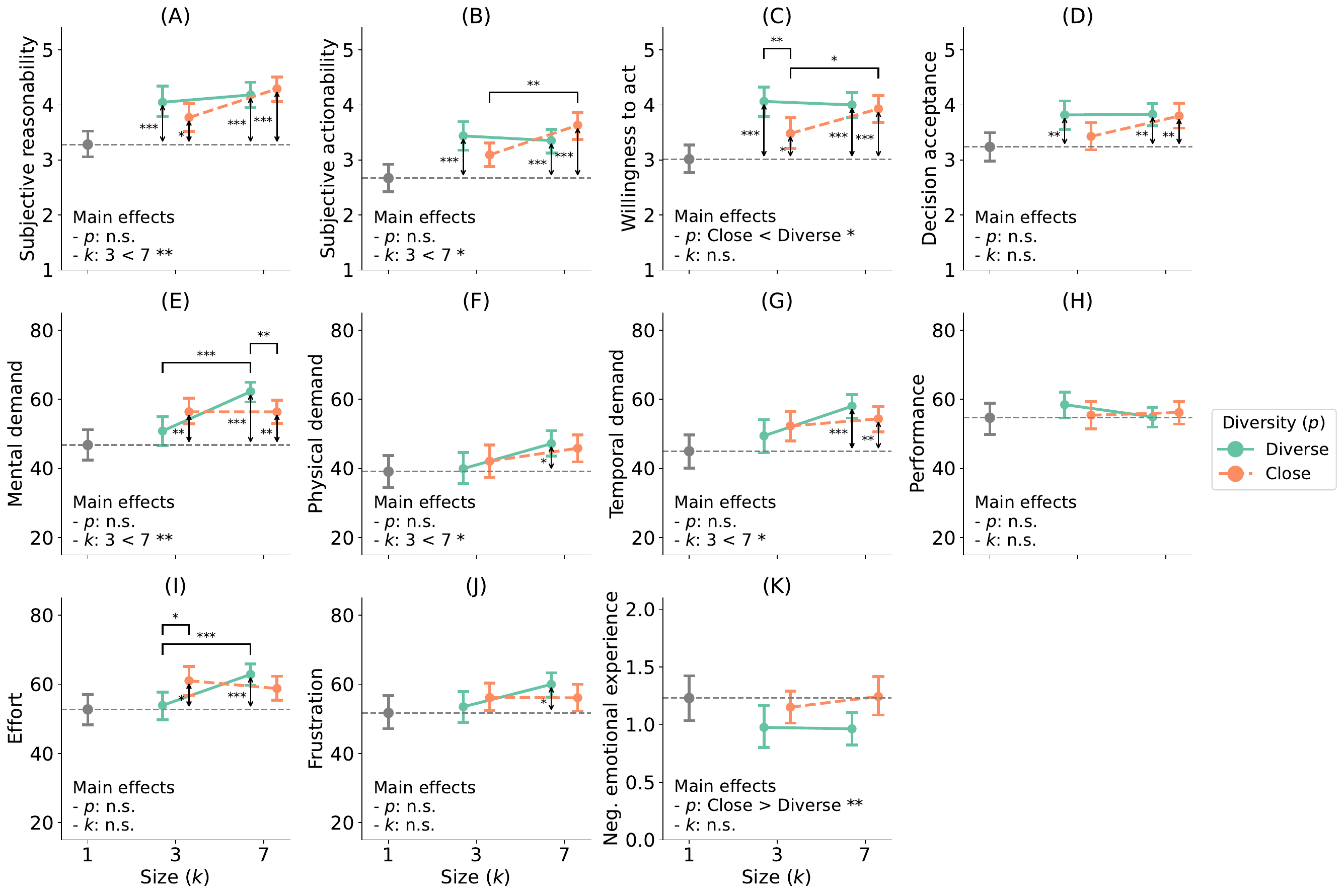}
    \caption{Psychological benefits (A)--(D) and costs (E)--(K). Here, (F) Physical demand, (G) Temporal demand, and (H) Performance are additionally reported, while the remaining outcomes are identical to those in the main paper. We report (i) two-way ANOVAs with recourse-set diversity $p$ and size $k$, followed by Tukey’s HSD post-hoc tests, and (ii) Dunnett’s one-to-many comparisons using Close–1 as the reference. All $p$-values are adjusted ($^*$: $p{<}0.05$, $^{**}$: $p{<}0.01$, $^{***}$: $p{<}0.001$). Main effects from the ANOVAs are reported in the lower-left of each panel. Tukey’s HSD results are indicated by umbrella symbols, and Dunnett’s comparisons are indicated by double-arrow symbols.}
    \label{supp_fig:tukey_all}
\end{figure}

Figure~\ref{supp_fig:tukey_all} summarizes how all outcome variables in our study vary as a function of recourse-set diversity and set size. 
In this supplementary material, we additionally report three outcomes: (F) Physical demand, (G) Temporal demand, and (H) Performance, while all other outcomes are identical to those in the main paper. 
For each outcome, we provide the results of two-way ANOVAs, Tukey's HSD post-hoc comparisons conducted only when the interaction is significant, and Dunnett's many-to-one comparisons using Close--1 as the reference condition, compiled in Tables~\ref{supp_tab:perceived_reasonability}--\ref{supp_tab:negative_emotional_experience}.

\begin{table}[t]
    \captionsetup{skip=5pt}
    \setlength{\tabcolsep}{0pt}
    \centering
    \small
    \begin{tabular}{lC{0.14\columnwidth}C{0.14\columnwidth}C{0.14\columnwidth}C{0.14\columnwidth}C{0.14\columnwidth}}
        \toprule
        \multicolumn{6}{l}{\textbf{Descriptive statistics}} \\
        \midrule
         & Close--1 & Close--3 & Diverse--3 & Close--7 & Diverse--7 \\
        \midrule
        Mean & 3.281 & 3.770 & 4.048 & 4.292 & 4.182 \\
        $N$  & 139 & 152 & 126 & 168 & 165 \\
        \midrule
        
        \addlinespace[1ex]
        \multicolumn{6}{l}{\textbf{Two-way ANOVA (Diversity $\times$ Size; excluding Close--1)}} \\
        \midrule
         & sum\_sq & df$_1$, df$_2$ & $F$ & $p_{\mathrm{adj}}$ & $\eta_p^2$ \\
        \midrule
        Diversity & \ \ 0.657 & 1, 607 & 0.297 & 0.586 & 0.000 \\
        Size & 17.358 & 1, 607 & 7.852 & 0.005 & 0.013 \\
        Diversity $\times$ Size & \ \ 5.667 & 1, 607 & 2.563 & 0.110 & 0.004 \\
        \midrule

        \addlinespace[1ex]
        \multicolumn{6}{l}{\textbf{Tukey's HSD post-hoc comparison (performed only if interaction is significant)}} \\
        \midrule
        Not performed.\\
        \midrule

        \addlinespace[1ex]
        \multicolumn{6}{l}{\textbf{Dunnett's multiple comparison (using Close--1 as reference)}} \\
        \midrule
        Contrast & $\Delta$ & 95\% CI & $t$ & $p_{\mathrm{adj}}$ & $g$ \\
        \midrule
        Close--3 $-$ Close--1   & 0.489 & [0.061, 0.918] & 2.787 & 0.019 & 0.317 \\
        Diverse--3 $-$ Close--1 & 0.767 & [0.318, 1.216] & 4.170 & $<$.001 & 0.502 \\
        Close--7 $-$ Close--1   & 1.011 & [0.593, 1.430] & 5.897 & $<$.001 & 0.687 \\
        Diverse--7 $-$ Close--1 & 0.901 & [0.481, 1.321] & 5.235 & $<$.001 & 0.596 \\
        \bottomrule
        
        \multicolumn{6}{l}{$SS$: sum of squares, df: degree of freedom, $F$: F statistics, $p_\mathrm{adj}$: adjusted p-value, $\eta_p^2$: partial eta squared}\\
        \multicolumn{6}{l}{$\Delta$: mean difference, CI: confidence intervals, $q$: studentized range statistic, $t$: t statistic, $g$: Hedges' g}
    \end{tabular}
     \caption{Perceived reasonability}\label{supp_tab:perceived_reasonability}
\end{table}

\begin{table}[t]
    \captionsetup{skip=5pt}
    \setlength{\tabcolsep}{0pt}
    \centering
    \small
    \begin{tabular}{lC{0.14\columnwidth}C{0.14\columnwidth}C{0.14\columnwidth}C{0.14\columnwidth}C{0.14\columnwidth}}
        \toprule
        \multicolumn{6}{l}{\textbf{Descriptive statistics}} \\
        \midrule
         & Close--1 & Close--3 & Diverse--3 & Close--7 & Diverse--7 \\
        \midrule
        Mean & 2.669 & 3.092 & 3.437 & 3.631 & 3.352 \\
        $N$  & 139 & 152 & 126 & 168 & 165 \\
        \midrule
        
        \addlinespace[1ex]
        \multicolumn{6}{l}{\textbf{Two-way ANOVA (Diversity $\times$ Size; excluding Close--1)}} \\
        \midrule
         & $SS$ & df$_1$, df$_2$ & $F$ & $p_{\mathrm{adj}}$ & $\eta_p^2$ \\
        \midrule
        Diversity & \ \ 0.001 & 1, 607 & 0.001 & 0.980 & 0.000 \\
        Size & \ \ 9.016 & 1, 607 & 4.041 & 0.045 & 0.007 \\
        Diversity $\times$ Size & 14.670 & 1, 607 & 6.575 & 0.011 & 0.011 \\
        \midrule

        \addlinespace[1ex]
        \multicolumn{6}{l}{\textbf{Tukey's HSD post-hoc comparison (performed only if interaction is significant)}} \\
        \midrule
        Contrast & $\Delta$ & 95\% CI & $q$ & $p_{\mathrm{adj}}$ & $g$ \\
        \midrule
        Close--3 $-$ Close--7     & $-$0.539 & [$-$0.877, $-$0.200] & 4.427 & 0.002 & $-$0.350 \\
        Diverse--3 $-$ Diverse--7 & \ \ \ 0.085 & [$-$0.251, \ \ \ 0.421] & 0.704 & 0.619 & \ \ \ 0.059 \\
        Close--3 $-$ Diverse--3   & $-$0.344 & [$-$0.695, \ \ \ 0.006] & 2.733 & 0.054 & $-$0.232 \\
        Close--7 $-$ Diverse--7   & \ \ \ 0.279 & [$-$0.045, \ \ \ 0.604] & 2.394 & 0.091 & \ \ \ 0.185 \\
        \midrule

        \addlinespace[1ex]
        \multicolumn{6}{l}{\textbf{Dunnett's multiple comparison (using Close--1 as reference)}} \\
        \midrule
        Contrast & $\Delta$ & 95\% CI & $t$ & $p_{\mathrm{adj}}$ & $g$ \\
        \midrule
        Close--3 $-$ Close--1   & 0.423 & [$-$0.002, 0.848] & 2.427 & 0.052 & 0.289 \\
        Diverse--3 $-$ Close--1 & 0.767 & [\ \ \ 0.322, 1.213] & 4.201 & $<$.001 & 0.522 \\
        Close--7 $-$ Close--1   & 0.962 & [\ \ \ 0.546, 1.378] & 5.649 & $<$.001 & 0.627 \\
        Diverse--7 $-$ Close--1 & 0.682 & [\ \ \ 0.265, 1.100] & 3.991 & $<$.001 & 0.477 \\
        \bottomrule
        
        \multicolumn{6}{l}{$SS$: sum of squares, df: degree of freedom, $F$: F statistic, $p_\mathrm{adj}$: adjusted p-value, $\eta_p^2$: partial eta squared}\\
        \multicolumn{6}{l}{$\Delta$: mean difference, CI: confidence interval, $q$: studentized range statistic, $t$: t statistic, $g$: Hedges' g}
    \end{tabular}
    \caption{Perceived actionability}\label{supp_tab:perceived_actionability}
\end{table}

\begin{table}[t]
    \captionsetup{skip=5pt}
    \setlength{\tabcolsep}{0pt}
    \centering
    \small
    \begin{tabular}{lC{0.14\columnwidth}C{0.14\columnwidth}C{0.14\columnwidth}C{0.14\columnwidth}C{0.14\columnwidth}}
        \toprule
        \multicolumn{6}{l}{\textbf{Descriptive statistics}} \\
        \midrule
         & Close--1 & Close--3 & Diverse--3 & Close--7 & Diverse--7 \\
        \midrule
        Mean & 3.014 & 3.480 & 4.063 & 3.929 & 4.000 \\
        $N$  & 139 & 152 & 126 & 168 & 165 \\
        \midrule
        
        \addlinespace[1ex]
        \multicolumn{6}{l}{\textbf{Two-way ANOVA (Diversity $\times$ Size; excluding Close--1)}} \\
        \midrule
         & $SS$ & df$_1$, df$_2$ & $F$ & $p_{\mathrm{adj}}$ & $\eta_p^2$ \\
        \midrule
        Diversity & 13.985 & 1, 607 & 5.672 & 0.018 & 0.009 \\
        Size & \ \ 6.452 & 1, 607 & 2.617 & 0.106 & 0.004 \\
        Diversity $\times$ Size & \ \ 9.874 & 1, 607 & 4.005 & 0.046 & 0.007 \\
        \midrule

        \addlinespace[1ex]
        \multicolumn{6}{l}{\textbf{Tukey's HSD post-hoc comparison (performed only if interaction is significant)}} \\
        \midrule
        Contrast & $\Delta$ & 95\% CI & $q$ & $p_{\mathrm{adj}}$ & $g$ \\
        \midrule
        Close--3 $-$ Close--7     & $-$0.448 & [$-$0.813, $-$0.084] & 3.422 & 0.016 & $-$0.270 \\
        Diverse--3 $-$ Diverse--7 & \ \ \ 0.064 & [$-$0.279, \ \ \ 0.406] & 0.516 & 0.716 & \ \ \ 0.043 \\
        Close--3 $-$ Diverse--3   & $-$0.583 & [$-$0.956, $-$0.210] & 4.351 & 0.002 & $-$0.370 \\
        Close--7 $-$ Diverse--7   & $-$0.071 & [$-$0.409, \ \ \ 0.266] & 0.588 & 0.678 & $-$0.045 \\
        \midrule

        \addlinespace[1ex]
        \multicolumn{6}{l}{\textbf{Dunnet's multiple comparison (using Close--1 as reference)}} \\
        \midrule
        Contrast & $\Delta$ & 95\% CI & $t$ & $p_{\mathrm{adj}}$ & $g$ \\
        \midrule
        Close--3 $-$ Close--1   & 0.466 & [0.017, 0.915] & 2.532 & 0.039 & 0.288 \\
        Diverse--3 $-$ Close--1 & 1.049 & [0.578, 1.520] & 5.441 & $<$.001 & 0.693 \\
        Close--7 $-$ Close--1   & 0.914 & [0.475, 1.353] & 5.086 & $<$.001 & 0.568 \\
        Diverse--7 $-$ Close--1 & 0.986 & [0.545, 1.426] & 5.461 & $<$.001 & 0.649 \\
        \bottomrule
        
        \multicolumn{6}{l}{$SS$: sum of squares, df: degree of freedom, $F$: F statistics, $p_\mathrm{adj}$: adjusted p-value, $\eta_p^2$: partial eta squared}\\
        \multicolumn{6}{l}{$\Delta$: mean difference, CI: confidence intervals, $q$: q statistics, $t$: t statistics, $g$: Hedges' g}
    \end{tabular}
    \caption{Willingness to act}\label{supp_tab:willingness_to_act}
\end{table}

\begin{table}[t]
    \captionsetup{skip=5pt}
    \setlength{\tabcolsep}{0pt}
    \centering
    \small
    \begin{tabular}{lC{0.14\columnwidth}C{0.14\columnwidth}C{0.14\columnwidth}C{0.14\columnwidth}C{0.14\columnwidth}}
        \toprule
        \multicolumn{6}{l}{\textbf{Descriptive statistics}} \\
        \midrule
         & Close--1 & Close--3 & Diverse--3 & Close--7 & Diverse--7 \\
        \midrule
        Mean & 3.237 & 3.428 & 3.817 & 3.798 & 3.830 \\
        $N$  & 139 & 152 & 126 & 168 & 165 \\
        \midrule
        
        \addlinespace[1ex]
        \multicolumn{6}{l}{\textbf{Two-way ANOVA (Diversity $\times$ Size; excluding Close--1)}} \\
        \midrule
         & $SS$ & df$_1$, df$_2$ & $F$ & $p_{\mathrm{adj}}$ & $\eta_p^2$ \\
        \midrule
        Diversity & 5.750 & 1, 607 & 2.581 & 0.109 & 0.004 \\
        Size & 6.128 & 1, 607 & 2.750 & 0.098 & 0.005 \\
        Diversity $\times$ Size & 4.808 & 1, 607 & 2.158 & 0.142 & 0.004 \\
        \midrule

        \addlinespace[1ex]
        \multicolumn{6}{l}{\textbf{Tukey's HSD post-hoc comparison (performed only if interaction is significant)}} \\
        \midrule
        Not performed. \\
        \midrule

        \addlinespace[1ex]
        \multicolumn{6}{l}{\textbf{Dunnet's multiple comparison (using Close--1 as reference)}} \\
        \midrule
        Contrast & $\Delta$ & 95\% CI & $t$ & $p_{\mathrm{adj}}$ & $g$ \\
        \midrule
        Close--3 $-$ Close--1   & 0.190 & [$-$0.242, 0.623] & 1.074 & 0.651 & 0.120 \\
        Diverse--3 $-$ Close--1 & 0.580 & [\ \ \ 0.127, 1.033] & 3.124 & 0.007 & 0.374 \\
        Close--7 $-$ Close--1   & 0.560 & [\ \ \ 0.138, 0.983] & 3.237 & 0.005 & 0.360 \\
        Diverse--7 $-$ Close--1 & 0.593 & [\ \ \ 0.169, 1.017] & 3.412 & 0.003 & 0.406 \\
        \bottomrule
        
        \multicolumn{6}{l}{$SS$: sum of squares, df: degree of freedom, $F$: F statistics, $p_\mathrm{adj}$: adjusted p-value, $\eta_p^2$: partial eta squared}\\
        \multicolumn{6}{l}{$\Delta$: mean difference, CI: confidence intervals, $q$: q statistics, $t$: t statistics, $g$: Hedges' g}
    \end{tabular}
    \caption{Decision acceptance}\label{supp_tab:decision_acceptance}
\end{table}

\begin{table}[t]
    \captionsetup{skip=5pt}
    \setlength{\tabcolsep}{0pt}
    \centering
    \small
    \begin{tabular}{lC{0.14\columnwidth}C{0.14\columnwidth}C{0.14\columnwidth}C{0.14\columnwidth}C{0.14\columnwidth}}
        \toprule
        \multicolumn{6}{l}{\textbf{Descriptive statistics}} \\
        \midrule
         & Close--1 & Close--3 & Diverse--3 & Close--7 & Diverse--7 \\
        \midrule
        Mean & 46.799 & 56.382 & 50.833 & 56.369 & 62.212 \\
        $N$  & 139 & 152 & 126 & 168 & 165 \\
        \midrule
        
        \addlinespace[1ex]
        \multicolumn{6}{l}{\textbf{Two-way ANOVA (Diversity $\times$ Size; excluding Close--1)}} \\
        \midrule
         & SS & df$_1$, df$_2$ & $F$ & $p_{\mathrm{adj}}$ & $\eta_p^2$ \\
        \midrule
        Diversity & \ \ \ \ 71.322 & 1, 607 & \ \ 0.148 & 0.701 & 0.000 \\
        Size & 4358.828 & 1, 607 & \ \ 9.019 & 0.003 & 0.015 \\
        Diversity $\times$ Size & 4891.435 & 1, 607 & 10.121 & 0.002 & 0.016 \\
        \midrule

        \addlinespace[1ex]
        \multicolumn{6}{l}{\textbf{Tukey's HSD post-hoc comparison (performed only if interaction is significant)}} \\
        \midrule
        Contrast & $\Delta$ & 95\% CI & $q$ & $p_{\mathrm{adj}}$ & $g$ \\
        \midrule
        Close--3 $-$ Close--7     & \ \ \ 0.013 & [\ \ $-$5.107, \ \ \ 5.132] & 0.007 & 0.996 & \ \ \ 0.001 \\
        Diverse--3 $-$ Diverse--7 & $-$11.379 & [$-$16.154, $-$6.603] & 6.632 & $<$.001 & $-$0.553 \\
        Close--3 $-$ Diverse--3   & \ \ \ 5.548 & [\ \ $-$0.087, \ 11.183] & 2.741 & 0.054 & \ \ \ 0.233 \\
        Close--7 $-$ Diverse--7   & $-$5.843 & [$-$10.238, $-$1.448] & 3.698 & 0.009 & $-$0.286 \\
        \midrule

        \addlinespace[1ex]
        \multicolumn{6}{l}{\textbf{Dunnet's multiple comparison (using Close--1 as reference)}} \\
        \midrule
        Contrast & $\Delta$ & 95\% CI & $t$ & $p_{\mathrm{adj}}$ & $g$ \\
        \midrule
        Close--3 $-$ Close--1   & \ \ 9.583 & [\ \ \ 3.033, 16.133] & 3.571 & 0.001 & 0.379 \\
        Diverse--3 $-$ Close--1 & \ \ 4.035 & [$-$2.831, 10.900] & 1.434 & 0.403 & 0.161 \\
        Close--7 $-$ Close--1   & \ \ 9.570 & [\ \ \ 3.171, 15.970] & 3.650 & 0.001 & 0.393 \\
        Diverse--7 $-$ Close--1 & 15.414 & [\ \ \ 8.988, 21.839] & 5.855 & $<$.001 & 0.690 \\
        \bottomrule
        
        \multicolumn{6}{l}{SS: sum of squares, df: degree of freedom, $F$: F statistics, $p_\mathrm{adj}$: adjusted p-value, $\eta_p^2$: partial eta squared}\\
        \multicolumn{6}{l}{$\Delta$: mean difference, CI: confidence intervals, $q$: q statistics, $t$: t statistics, $g$: Hedges' g}
    \end{tabular}
    \caption{Mental demand}\label{supp_tab:mental_demand}
\end{table}

\begin{table}[t]
    \captionsetup{skip=5pt}
    \setlength{\tabcolsep}{0pt}
    \centering
    \small
    \begin{tabular}{lC{0.14\columnwidth}C{0.14\columnwidth}C{0.14\columnwidth}C{0.14\columnwidth}C{0.14\columnwidth}}
        \toprule
        \multicolumn{6}{l}{\textbf{Descriptive statistics}} \\
        \midrule
         & Close--1 & Close--3 & Diverse--3 & Close--7 & Diverse--7 \\
        \midrule
        Mean & 39.101 & 42.105 & 39.960 & 45.863 & 47.182 \\
        $N$  & 139 & 152 & 126 & 168 & 165 \\
        \midrule
        
        \addlinespace[1ex]
        \multicolumn{6}{l}{\textbf{Two-way ANOVA (Diversity $\times$ Size; excluding Close--1)}} \\
        \midrule
         & $SS$ & df$_1$, df$_2$ & $F$ & $p_{\mathrm{adj}}$ & $\eta_p^2$ \\
        \midrule
        Diversity & \ \ \ \ \ \ 9.489 & 1, 607 & 0.013 & 0.909 & 0.000 \\
        Size & 4400.422 & 1, 607 & 6.093 & 0.014 & 0.010 \\
        Diversity $\times$ Size & \ \ 452.232 & 1, 607 & 0.626 & 0.429 & 0.001 \\
        \midrule

        \addlinespace[1ex]
        \multicolumn{6}{l}{\textbf{Tukey's HSD post-hoc comparison (performed only if interaction is significant)}} \\
        \midrule
        Not performed. \\
        \midrule

        \addlinespace[1ex]
        \multicolumn{6}{l}{\textbf{Dunnet's multiple comparison (using Close--1 as reference)}} \\
        \midrule
        Contrast & $\Delta$ & 95\% CI & $t$ & $p_{\mathrm{adj}}$ & $g$ \\
        \midrule
        Close--3 $-$ Close--1   & 3.005 & [$-$4.755, 10.764] & 0.945 & 0.742 & 0.105 \\
        Diverse--3 $-$ Close--1 & 0.860 & [$-$7.273, \ \ 8.992] & 0.258 & 0.997 & 0.031 \\
        Close--7 $-$ Close--1   & 6.762 & [$-$0.818, 14.343] & 2.177 & 0.096 & 0.249 \\
        Diverse--7 $-$ Close--1 & 8.081 & [\ \ \ 0.469, 15.693] & 2.591 & 0.034 & 0.306 \\
        \bottomrule
        
        \multicolumn{6}{l}{$SS$: sum of squares, df: degree of freedom, $F$: F statistics, $p_\mathrm{adj}$: adjusted p-value, $\eta_p^2$: partial eta squared}\\
        \multicolumn{6}{l}{$\Delta$: mean difference, CI: confidential intervals, $q$: q statistics, $t$: t statistics, $g$: Hedges' g}
    \end{tabular}
    \caption{Physical demand}\label{supp_tab:physical_demand}
\end{table}

\begin{table}[t]
    \captionsetup{skip=5pt}
    \setlength{\tabcolsep}{0pt}
    \centering
    \small
    \begin{tabular}{lC{0.14\columnwidth}C{0.14\columnwidth}C{0.14\columnwidth}C{0.14\columnwidth}C{0.14\columnwidth}}
        \toprule
        \multicolumn{6}{l}{\textbf{Descriptive statistics}} \\
        \midrule
         & Close--1 & Close--3 & Diverse--3 & Close--7 & Diverse--7 \\
        \midrule
        Mean & 45.000 & 52.303 & 49.444 & 54.345 & 58.000 \\
        $N$  & 139 & 152 & 126 & 168 & 165 \\
        \midrule
        
        \addlinespace[1ex]
        \multicolumn{6}{l}{\textbf{Two-way ANOVA (Diversity $\times$ Size; excluding Close--1)}} \\
        \midrule
         & $SS$ & df$_1$, df$_2$ & $F$ & $p_{\mathrm{adj}}$ & $\eta_p^2$ \\
        \midrule
        Diversity & \ \ \ \ 75.717 & 1, 607 & 0.117 & 0.733 & 0.000 \\
        Size & 3963.436 & 1, 607 & 6.120 & 0.014 & 0.010 \\
        Diversity $\times$ Size & 1598.982 & 1, 607 & 2.469 & 0.117 & 0.004 \\
        \midrule

        \addlinespace[1ex]
        \multicolumn{6}{l}{\textbf{Tukey's HSD post-hoc comparison (performed only if interaction is significant)}} \\
        \midrule
        Not performed. \\
        \midrule

        \addlinespace[1ex]
        \multicolumn{6}{l}{\textbf{Dunnet's multiple comparison (using Close--1 as reference)}} \\
        \midrule
        Contrast & $\Delta$ & 95\% CI & $t$ & $p_{\mathrm{adj}}$ & $g$ \\
        \midrule
        Close--3 $-$ Close--1   & \ \ 7.303 & [$-$0.194, 14.800] & 2.378 & 0.059 & 0.258 \\
        Diverse--3 $-$ Close--1 & \ \ 4.444 & [$-$3.413, 12.302] & 1.381 & 0.437 & 0.155 \\
        Close--7 $-$ Close--1   & \ \ 9.345 & [\ \ \ 2.021, 16.670] & 3.114 & 0.007 & 0.352 \\
        Diverse--7 $-$ Close--1 & 13.000 & [\ \ \ 5.646, 20.354] & 4.315 & $<$.001 & 0.501 \\
        \bottomrule
        
        \multicolumn{6}{l}{$SS$: sum of squares, df: degree of freedom, $F$: F statistics, $p_\mathrm{adj}$: adjusted p-value, $\eta_p^2$: partial eta squared}\\
        \multicolumn{6}{l}{$\Delta$: mean difference, CI: confidential intervals, $q$: q statistics, $t$: t statistics, $g$: Hedges' g}
    \end{tabular}
    \caption{Temporal demand}\label{supp_tab:temporal_demand}
\end{table}

\begin{table}[t]
    \captionsetup{skip=5pt}
    \setlength{\tabcolsep}{0pt}
    \centering
    \small
    \begin{tabular}{lC{0.14\columnwidth}C{0.14\columnwidth}C{0.14\columnwidth}C{0.14\columnwidth}C{0.14\columnwidth}}
        \toprule
        \multicolumn{6}{l}{\textbf{Descriptive statistics}} \\
        \midrule
         & Close--1 & Close--3 & Diverse--3 & Close--7 & Diverse--7 \\
        \midrule
        Mean & 54.676 & 55.395 & 58.452 & 56.190 & 54.909 \\
        $N$  & 139 & 152 & 126 & 168 & 165 \\
        \midrule
        
        \addlinespace[1ex]
        \multicolumn{6}{l}{\textbf{Two-way ANOVA (Diversity $\times$ Size; excluding Close--1)}} \\
        \midrule
         & $SS$ & df$_1$, df$_2$ & $F$ & $p_{\mathrm{adj}}$ & $\eta_p^2$ \\
        \midrule
        Diversity & \ \ 71.068 & 1, 607 & 0.155 & 0.694 & 0.000 \\
        Size & 237.795 & 1, 607 & 0.518 & 0.472 & 0.001 \\
        Diversity $\times$ Size & 709.698 & 1, 607 & 1.545 & 0.214 & 0.003 \\
        \midrule

        \addlinespace[1ex]
        \multicolumn{6}{l}{\textbf{Tukey's HSD post-hoc comparison (performed only if interaction is significant)}} \\
        \midrule
        Not performed. \\
        \midrule

        \addlinespace[1ex]
        \multicolumn{6}{l}{\textbf{Dunnet's multiple comparison (using Close--1 as reference)}} \\
        \midrule
        Contrast & $\Delta$ & 95\% CI & $t$ & $p_{\mathrm{adj}}$ & $g$ \\
        \midrule
        Close--3 $-$ Close--1   & 0.718 & [$-$5.678, \ \ 7.115] & 0.274 & 0.996 & 0.029 \\
        Diverse--3 $-$ Close--1 & 3.776 & [$-$2.928, 10.480] & 1.375 & 0.441 & 0.159 \\
        Close--7 $-$ Close--1   & 1.514 & [$-$4.735, \ \ 7.763] & 0.591 & 0.935 & 0.063 \\
        Diverse--7 $-$ Close--1 & 0.233 & [$-$6.042, \ \ 6.508] & 0.091 & 1.000 & 0.010 \\
        \bottomrule
        
        \multicolumn{6}{l}{$SS$: sum of squares, df: degree of freedom, $F$: F statistics, $p_\mathrm{adj}$: adjusted p-value, $\eta_p^2$: partial eta squared}\\
        \multicolumn{6}{l}{$\Delta$: mean difference, CI: confidential intervals, $q$: q statistics, $t$: t statistics, $g$: Hedges' g}
    \end{tabular}
    \caption{Performance}\label{supp_tab:performance}
\end{table}

\begin{table}[t]
    \captionsetup{skip=5pt}
    \setlength{\tabcolsep}{0pt}
    \centering
    \small
    \begin{tabular}{lC{0.14\columnwidth}C{0.14\columnwidth}C{0.14\columnwidth}C{0.14\columnwidth}C{0.14\columnwidth}}
        \toprule
        \multicolumn{6}{l}{\textbf{Descriptive statistics}} \\
        \midrule
         & Close--1 & Close--3 & Diverse--3 & Close--7 & Diverse--7 \\
        \midrule
        Mean & 52.734 & 61.020 & 53.889 & 58.780 & 62.848 \\
        $N$  & 139 & 152 & 126 & 168 & 165 \\
        \midrule
        
        \addlinespace[1ex]
        \multicolumn{6}{l}{\textbf{Two-way ANOVA (Diversity $\times$ Size; excluding Close--1)}} \\
        \midrule
         & $SS$ & df$_1$, df$_2$ & $F$ & $p_{\mathrm{adj}}$ & $\eta_p^2$ \\
        \midrule
        Diversity & \ \ 152.996 & 1, 607 & 0.284 & 0.594 & 0.000 \\
        Size & 1407.313 & 1, 607 & 2.611 & 0.107 & 0.004 \\
        Diversity $\times$ Size & 4728.148 & 1, 607 & 8.773 & 0.003 & 0.014 \\
        \midrule

        \addlinespace[1ex]
        \multicolumn{6}{l}{\textbf{Tukey's HSD post-hoc comparison (performed only if interaction is significant)}} \\
        \midrule
        Contrast & $\Delta$ & 95\% CI & $q$ & $p_{\mathrm{adj}}$ & $g$ \\
        \midrule
        Close--3 $-$ Close--7     & \ \ \ 2.240 & [\ \ $-$3.110, \ \ \ 7.590] & 1.165 & 0.411 & \ \ \ 0.092 \\
        Diverse--3 $-$ Diverse--7 & $-$8.960 & [$-$14.076, $-$3.844] & 4.875 & $<$.001 & $-$0.407 \\
        Close--3 $-$ Diverse--3   & \ \ \ 7.131 & [\ \ \ \ \ 1.252, \ 13.010] & 3.377 & 0.018 & \ \ \ 0.287 \\
        Close--7 $-$ Diverse--7   & $-$4.069 & [\ \ $-$8.773, \ \ \ 0.636] & 2.406 & 0.090 & $-$0.186 \\
        \midrule

        \addlinespace[1ex]
        \multicolumn{6}{l}{\textbf{Dunnet's multiple comparison (using Close--1 as reference)}} \\
        \midrule
        Contrast & $\Delta$ & 95\% CI & $t$ & $p_{\mathrm{adj}}$ & $g$ \\
        \midrule
        Close--3 $-$ Close--1   & \ \ 8.286 & [\ \ \ 1.459, 15.113] & 2.962 & 0.011 & 0.318 \\
        Diverse--3 $-$ Close--1 & \ \ 1.155 & [$-$6.001, \ \ 8.311] & 0.394 & 0.984 & 0.046 \\
        Close--7 $-$ Close--1   & \ \ 6.046 & [$-$0.624, 12.716] & 2.212 & 0.088 & 0.245 \\
        Diverse--7 $-$ Close--1 & 10.115 & [\ \ \ 3.417, 16.812] & 3.686 & $<$.001 & 0.432 \\
        \bottomrule
        
        \multicolumn{6}{l}{$SS$: sum of squares, df: degree of freedom, $F$: F statistics, $p_\mathrm{adj}$: adjusted p-value, $\eta_p^2$: partial eta squared}\\
        \multicolumn{6}{l}{$\Delta$: mean difference, CI: confidential intervals, $q$: q statistics, $t$: t statistics, $g$: Hedges' g}
    \end{tabular}
    \caption{Effort}\label{supp_tab:effort}
\end{table}

\begin{table}[t]
    \captionsetup{skip=5pt}
    \setlength{\tabcolsep}{0pt}
    \centering
    \small
    \begin{tabular}{lC{0.14\columnwidth}C{0.14\columnwidth}C{0.14\columnwidth}C{0.14\columnwidth}C{0.14\columnwidth}}
        \toprule
        \multicolumn{6}{l}{\textbf{Descriptive statistics}} \\
        \midrule
         & Close--1 & Close--3 & Diverse--3 & Close--7 & Diverse--7 \\
        \midrule
        Mean & 51.727 & 56.151 & 53.492 & 56.101 & 59.970 \\
        $N$  & 139 & 152 & 126 & 168 & 165 \\
        \midrule
        
        \addlinespace[1ex]
        \multicolumn{6}{l}{\textbf{Two-way ANOVA (Diversity $\times$ Size; excluding Close--1)}} \\
        \midrule
         & $SS$ & df$_1$, df$_2$ & $F$ & $p_{\mathrm{adj}}$ & $\eta_p^2$ \\
        \midrule
        Diversity & \ \ 126.680 & 1, 607 & 0.202 & 0.653 & 0.000 \\
        Size & 1391.680 & 1, 607 & 2.216 & 0.137 & 0.004 \\
        Diversity $\times$ Size & 1606.262 & 1, 607 & 2.558 & 0.110 & 0.004 \\
        \midrule

        \addlinespace[1ex]
        \multicolumn{6}{l}{\textbf{Tukey's HSD post-hoc comparison (performed only if interaction is significant)}} \\
        \midrule
        Not performed. \\
        \midrule

        \addlinespace[1ex]
        \multicolumn{6}{l}{\textbf{Dunnet's multiple comparison (using Close--1 as reference)}} \\
        \midrule
        Contrast & $\Delta$ & 95\% CI & $t$ & $p_{\mathrm{adj}}$ & $g$ \\
        \midrule
        Close--3 $-$ Close--1   & 4.425 & [$-$2.946, 11.795] & 1.465 & 0.384 & 0.163 \\
        Diverse--3 $-$ Close--1 & 1.765 & [$-$5.960, \ \ 9.491] & 0.558 & 0.947 & 0.064 \\
        Close--7 $-$ Close--1   & 4.375 & [$-$2.826, 11.576] & 1.483 & 0.373 & 0.162 \\
        Diverse--7 $-$ Close--1 & 8.243 & [\ \ \ 1.013, 15.474] & 2.783 & 0.020 & 0.319 \\
        \bottomrule
        
        \multicolumn{6}{l}{$SS$: sum of squares, df: degree of freedom, $F$: F statistics, $p_\mathrm{adj}$: adjusted p-value, $\eta_p^2$: partial eta squared}\\
        \multicolumn{6}{l}{$\Delta$: mean difference, CI: confidential intervals, $q$: q statistics, $t$: t statistics, $g$: Hedges' g}
    \end{tabular}
    \caption{Frustration}\label{supp_tab:frustration}
\end{table}

\begin{table}[t]
    \captionsetup{skip=5pt}
    \setlength{\tabcolsep}{0pt}
    \centering
    \small
    \begin{tabular}{lC{0.14\columnwidth}C{0.14\columnwidth}C{0.14\columnwidth}C{0.14\columnwidth}C{0.14\columnwidth}}
        \toprule
        \multicolumn{6}{l}{\textbf{Descriptive statistics}} \\
        \midrule
         & Close--1 & Close--3 & Diverse--3 & Close--7 & Diverse--7 \\
        \midrule
        Mean & 1.230 & 1.151 & 0.976 & 1.244 & 0.964 \\
        $N$  & 139 & 152 & 126 & 168 & 165 \\
        \midrule
        
        \addlinespace[1ex]
        \multicolumn{6}{l}{\textbf{Two-way ANOVA (Diversity $\times$ Size; excluding Close--1)}} \\
        \midrule
         & $SS$ & df$_1$, df$_2$ & $F$ & $p_{\mathrm{adj}}$ & $\eta_p^2$ \\
        \midrule
        Diversity & 8.240 & 1, 607 & 7.949 & 0.005 & 0.013 \\
        Size & 0.280 & 1, 607 & 0.270 & 0.604 & 0.000 \\
        Diversity $\times$ Size & 0.418 & 1, 607 & 0.403 & 0.526 & 0.001 \\
        \midrule

        \addlinespace[1ex]
        \multicolumn{6}{l}{\textbf{Tukey's HSD post-hoc comparison (performed only if interaction is significant)}} \\
        \midrule
        Not performed. \\
        \midrule

        \addlinespace[1ex]
        \multicolumn{6}{l}{\textbf{Dunnet's multiple comparison (using Close--1 as reference)}} \\
        \midrule
        Contrast & $\Delta$ & 95\% CI & $t$ & $p_{\mathrm{adj}}$ & $g$ \\
        \midrule
        Close--3 $-$ Close--1   & $-$0.079 & [$-$0.383, 0.225] & $-$0.634 & 0.919 & $-$0.071 \\
        Diverse--3 $-$ Close--1 & $-$0.254 & [$-$0.572, 0.064] & $-$1.947 & 0.158 & $-$0.219 \\
        Close--7 $-$ Close--1   & \ \ \ 0.014 & [$-$0.283, 0.311] & \ \ \ 0.114 & 1.000 & \ \ \ 0.012 \\
        Diverse--7 $-$ Close--1 & $-$0.267 & [$-$0.565, 0.031] & $-$2.183 & 0.094 & $-$0.246 \\
        \bottomrule
        
        \multicolumn{6}{l}{$SS$: sum of squares, df: degree of freedom, $F$: F statistics, $p_\mathrm{adj}$: adjusted p-value, $\eta_p^2$: partial eta squared}\\
        \multicolumn{6}{l}{$\Delta$: mean difference, CI: confidential intervals, $q$: q statistics, $t$: t statistics, $g$: Hedges' g}
    \end{tabular}
    \caption{Negative emotional experience}\label{supp_tab:negative_emotional_experience}
\end{table}

\FloatBarrier
\section{Qualitative Analysis and Results}\label{supp_subsec:qualitative_full}
To complement the quantitative results, we analyzed participants' open-ended justifications for subjective reasonability, subjective actionability, and willingness to act.  
Coding was conducted separately for each.  
One author first developed the codebook through iterative review of the dataset.  
To assess validity, an external collaborator independently coded a stratified random subsample of 75 responses per outcome measure (10\% of the data, stratified by condition).
Substantial agreement~\cite{Landis1977} was achieved between the coders (Cohen's kappa~\cite{Cohen1960} $\kappa=0.64$ for subjective reasonability, $0.63$ for subjective actionability, and $0.61$ for willingness to act).
The remaining responses were coded by the author, and codes were refined iteratively across outcomes and synthesized into overarching themes.
We show the codebook in Table~\ref{supp_tab:codebook} and compile the code distributions in Table~\ref{supp_tab:theme_code}.

\setlength{\tabcolsep}{3pt}
\renewcommand{\arraystretch}{1.04}
\newcolumntype{S}[1]{>{\small\raggedright\arraybackslash}p{#1}}
\begin{longtable}
{@{}S{0.12\columnwidth}S{0.01\columnwidth}S{0.19\columnwidth}S{0.64\columnwidth}@{}}
\toprule
\textbf{Theme} & \textbf{\#} & \textbf{Code} & \textbf{Definition \& Examples} \\
\midrule
\endfirsthead
\toprule
\textbf{Theme} & \textbf{\#} & \textbf{Code} & \textbf{Definition \& Examples} \\
\midrule
\endhead
\midrule
\multicolumn{4}{r}{\footnotesize(To be continued)}\\
\caption{Codebook: themes, codes, definitions, and example quotes}\label{supp_tab:codebook}\\
\endfoot
\bottomrule
\multicolumn{4}{r}{\footnotesize(The end of the table)}\\
\caption{Codebook: themes, codes, definitions, and example quotes (continued)}\\
\endlastfoot

\multicolumn{4}{@{}l}{\small\textbf{Subjective Reasonability}}\\
Feasibility \& Practicality & 1 & Likelihood of Feasibility
& Evaluations that the recourse action plan is possible, impossible, unrealistic, or too trivial; also includes remarks that only few people would qualify or only a small fraction could meet it; e.g., P646: ``\textit{It would be possible if I try.}''; P212: ``\textit{Practically impossible.}''; P911: ``\textit{I think only a small minority could satisfy these conditions.}'' \\

Personal Values \& Life Fit & 2 & Self-reflection
& Reflections on how the plan fits or does not fit one's actual circumstances, sometimes noting resignation; e.g., P431: ``\textit{I accepted it because I actually have few years of tenure.}''; P812: ``\textit{Given my current situation it's impossible, so it can't be helped.}'' \\

& 3 & Resigned Acceptance
& Reluctant acceptance arising from being in a disadvantaged position as the applicant or from a high bar; e.g., P731: ``\textit{If that's the reason for rejection, I have no choice but to accept it.}'' \\

Accountability & 4 & Appropriateness
& Judgments about whether the presented features or concepts are appropriate/inappropriate or important/unnecessary as evaluation criteria; includes pointing out criteria beyond one's control as inappropriate; e.g., P1788: ``\textit{Why would management experience be necessary?}''; P842: ``\textit{In the end, income is what matters most.}'' \\

& 5 & Lack of Clarity
& States not knowing which criteria were used, or that explanations were insufficient/unclear; includes dissatisfaction that necessary information was not provided; e.g., P309: ``\textit{I can't tell which factor is the reason.}''; P117: ``\textit{There's no information on income or savings.}'' \\

& 6 & Uncertainty and Risk
& Notes that the proposed actions have uncertain effects, lack guarantees, or carry risks, and/or expresses anxiety about that; e.g., P538: ``\textit{There are many sources of anxiety.}''; P623: ``\textit{Having a short tenure is a risk.}'' \\

& 7 & Unfairness
& Perceives the criteria or conditions as disadvantageous, unfair, or discriminatory; includes expressions of subjective unfairness or dissatisfaction; e.g., P901: ``\textit{It's unfair that renting puts you at a disadvantage.}''; P765: ``\textit{I think it's discriminatory.}'' \\

& 8 & Distrust of AI
& Distrust or aversion toward the AI/algorithmic evaluation process itself; e.g., P958: ``\textit{I dislike AI making the judgment.}''; P468: ``\textit{I feel mechanically discriminated against.}'' \\

\addlinespace
\multicolumn{4}{@{}l}{\small\textbf{Subjective Actionability}}\\
Effort \& Environmental Constraints & 1 & External Constraints
& Reasons based on factors beyond personal discretion—things that cannot be changed by one's own effort or will (e.g., company rules, social systems, others' decisions); e.g., P73: "I can't decide this on my own."; P74: "I can't raise my position by my own power alone."; P101: "My own effort alone won't change this." \\

& 2 & Time and Money
& Cites a large time or financial burden required to carry it out; e.g., P19: ``\textit{It requires achievements and time.}''; P152: ``\textit{Because it would take more than three years to reach the goal.}''; P157: ``\textit{It would take sustained effort, with results only after several years.}'' \\

& 3 & Manageability
& Sees it as manageable—requiring little effort, low burden, and within a doable range; e.g., P10: ``\textit{An increase of about two hours of work time wouldn't affect me much.''}; P11: ``\textit{Because it is feasible.}''; P49: ``\textit{Because it's just extending my working hours.}'' \\

& 4 & Skills and Experience
& Reasons related to learning/skill acquisition, credentialing, or lack of experience; e.g., P82: ``\textit{I would need the ability to reduce hours while keeping workload the same and earning the same pay.}''; P118: ``\textit{Can't change your educational background.}''; P211: ``\textit{I have limited work experience.}'' \\

& 5 & Social Status
& Cites occupational or social status factors (position, income, title, social credit); e.g., P67: ``\textit{Because I have to raise my job position.}''; P123: ``\textit{Job rank doesn't go up that easily.}''; P441: ``\textit{Because I would have to become the company president.}'' \\

Feasibility \& Practicality & 6 & Inconsistency
& Points out internal inconsistencies or contradictions within the recourse proposals; e.g., P130: ``\textit{Even if I switch to a public agency, my years of tenure would reset.}''; P496: ``\textit{To obtain a graduate degree I'd likely have to quit my job; increasing daily working hours on top of that seems contradictory.}'' \\

& 7 & General Difficulty
& General mentions of ``difficult/easy'' that don't fit other categories; e.g., P55: ``\textit{The barrier to changing jobs is high.}''; P93: ``\textit{There are things I can't do.}''; P168: ``\textit{Because I feel the requirements are a bit hard to achieve.}'' \\

Personal Velues \& Life Fit & 8 & Psychological Resistance
& Personal aversion or reluctance such as ``don't want to,'' ``not inclined,'' or ``feel resistance''; e.g., P48: ``\textit{The increase in working hours ...}''; P114: ``\textit{It's exhausting.}'' \\

& 9 & Impacts on/of Lifestyle
& Judges executability in light of one's lifestyle or life circumstances; e.g., P17: ``\textit{Because it's close to my real self.}''; P42: ``\textit{I've been thinking it's time to move and was considering buying a home.}''; P60: ``\textit{I don't want to change my life that much just to take out a loan.}'' \\

Accountability & 10 & Uncertainty and Risk
& Notes that the proposed actions have uncertain effects, lack guarantees, or carry risks, and/or expresses anxiety about that; e.g., P44: ``\textit{I don't know whether changing jobs would increase my income.}''; P138: ``\textit{I don't know if I can get promoted.}''; P153: ``\textit{Promotion is not in prospect.}'' \\

& 11 & Aversion to AI
& Based on distrust or aversion toward AI itself; e.g., P839: ``\textit{It feels like AI is belittling me--it's unpleasant, so I won't engage with it.}''; P2004: ``\textit{I doubt AI is reliable in the first place. It seems untenable.}'' \\

\addlinespace
\multicolumn{4}{@{}l}{\small\textbf{Willingness to act}}\\
Effort \& Environmental Constraints & 1 & External Constraints
& Reasons based on factors beyond personal discretion—things that cannot be changed by one's own effort or will (e.g., company rules, social systems, others' decisions); e.g., P286: ``\textit{I'd do it if I could decide myself, but I don't have the authority.}''; P1825: ``\textit{Changing position is heteronomous--it's not something I can do proactively myself.}''; P2668: ``\textit{It's beyond my control.}'' \\

& 2 & Degree of Effort
& Size of the burdens involved—time, hassle, exertion, or monetary cost; e.g., P44: ``\textit{Because I think I can get promoted and get a raise through my own effort.}''; P306: ``\textit{Because the burden would increase.}''; P1341: ``\textit{It might take some time, but it can be done without difficulty.}''; P1398: ``\textit{I felt it would not require that much effort.}'' \\

Feasibility \& Practicality & 3 & Feasibility Judgment
& Judgments of whether it can/can't be done, how difficult it is, or how certain it is; e.g., P168: ``\textit{Because I feel the action plan to meet the conditions is a bit difficult.}''; P670: ``\textit{Because it is the easiest.}''; P1901: ``\textit{Because it's not completely impossible.}'' \\

& 4 & Gains
& External benefits from acting (income, promotion, stability, health), or the lack thereof; e.g., P372: ``\textit{Because owning a home becomes an asset.}''; P1352: ``\textit{Because it would build my experience and be efficient.}''; P1624: ``\textit{Because building my career in Tokyo would give me social credit.}'' \\

Personal Velues \& Life Fit & 5 & Psychological Resistance
& Personal aversion or reluctance such as ``don't want to,'' ``not inclined,'' or ``feel resistance''; e.g., P789: ``\textit{Because it's a hassle.}''; P1307: ``\textit{I don't want to go to Tokyo and I don't want to work outside.}''; P1608: ``\textit{Because I don't want to increase my working hours.}'' \\

& 6 & Motives and Values
& Reasons grounded in personal desires, preferences, lifestyle, or sense of achievement; e.g., P547: ``\textit{Because I intend to do what I can.}''; P1178: ``\textit{I want to stay employed as long as possible.}''; P2637: ``\textit{I want to shorten my working hours and increase my private time.}'' \\

& 7 & Unnecessity
& Comments on the plan's necessity, meaning, prospects, or persuasiveness, or (non-)acceptance of the criteria; includes reluctant compliance; e.g., P68: ``\textit{Because I already own a home and don't feel a need to move.}''; P390: ``\textit{I can accept it.}''; P2355: ``\textit{It has future potential.}'' \\

Accountability & 8 & Uncertainty and Risk
& Notes that the proposed actions have uncertain effects, lack guarantees, or carry risks, and/or expresses anxiety about that; e.g., P1315: ``\textit{Because there's no guarantee of achieving it even with time.}''; P1941: ``\textit{The risk is too high.}''; P1989: ``\textit{There is no prospect of my position rising.}'' \\

& 9 & Unfairness
& Reasons grounded in social justice, fairness, or ethical norms, or in trust/distrust toward AI itself; e.g., P755: ``\textit{I can't trust AI that limits positions.}''; P2162: ``\textit{This is a very delicate matter.}'' \\

\end{longtable}
\begin{table}[t]
    \setlength{\tabcolsep}{3pt}
    \small
    \centering
    \begin{tabular}{lrlrrrrr}
        \toprule
         & & & \multicolumn{1}{c}{N=1} & \multicolumn{2}{c}{N=3} & \multicolumn{2}{c}{N=7} \\
        \cmidrule(lr){4-4} \cmidrule(lr){5-6} \cmidrule(lr){7-8}
        Theme & \multicolumn{2}{l}{Code} & \multicolumn{1}{c}{Close} & \multicolumn{1}{c}{Close} & \multicolumn{1}{c}{Diverse} & \multicolumn{1}{c}{Close} & \multicolumn{1}{c}{Diverse} \\
        \midrule
        \multicolumn{3}{l}{\textbf{Subjective Reasonability (SR)}}\\
        \ \ Feasibility \& Practicality & 1. & Likelihood of Feasibility & 13 (11.4\%) & 21 (16.8\%) & 17 (18.3\%) & 17 (13.1\%) & 18 (14.0\%) \\
        \ \ Personal Values \& Life Fit & 2. & Self-reflection & 12 (10.5\%) & 14 (11.2\%) & 16 (17.2\%) & 30 (23.1\%) & 31 (24.0\%) \\
         & 3. & Resigned Acceptance & 9 (7.9\%) & 7 (5.6\%) & 7 (7.5\%) & 9 (6.9\%) & 5 (3.9\%) \\
        \ \ Accountability & 4. & Appropriateness & 49 (43.0\%) & 51 (40.8\%) & 36 (38.7\%) & 51 (39.2\%) & 50 (38.8\%) \\
         & 5. & Lack of Clarity & 15 (13.2\%) & 11 (8.8\%) & 5 (5.4\%) & 9 (6.9\%) & 10 (7.8\%) \\
         & 6. & Uncertainty and Risk & 2 (1.8\%) & 7 (5.6\%) & 3 (3.2\%) & 4 (3.1\%) & 2 (1.6\%) \\
         & 7. & Unfairness & 4 (3.5\%) & 6 (4.8\%) & 3 (3.2\%) & 4 (3.1\%) & 6 (4.7\%) \\
         & 8. & Distrust of AI & 3 (2.6\%) & 1 (0.8\%) & 0 (0.0\%) & 0 (0.0\%) & 0 (0.0\%) \\
        \ \ Others &  &  & 7 (6.1\%) & 7 (5.6\%) & 6 (6.5\%) & 6 (4.6\%) & 7 (5.4\%) \\
        \ \ Total &  &  & 114 & 125 & 93 & 130 & 129 \\
        \midrule
        \multicolumn{3}{l}{\textbf{Subjective Actionability (SA)}}\\
        \ \ Effort \& Environmental Constraints & 1. & External Constraints & 27 (20.8\%) & 16 (11.9\%) & 13 (12.7\%) & 14 (9.9\%) & 13 (9.7\%) \\
         & 2. & Time and Money & 26 (20.0\%) & 13 (9.7\%) & 11 (10.8\%) & 16 (11.3\%) & 16 (11.9\%) \\
         & 3. & Manageability & 22 (16.9\%) & 26 (19.4\%) & 25 (24.5\%) & 43 (30.3\%) & 29 (21.6\%) \\
         & 4. & Skills and Experience & 14 (10.8\%) & 14 (10.4\%) & 6 (5.9\%) & 14 (9.9\%) & 9 (6.7\%) \\
         & 5. & Social Status & 8 (6.2\%) & 15 (11.2\%) & 6 (5.9\%) & 14 (9.9\%) & 16 (11.9\%) \\
        \ \ Feasibility \& Practicality & 6. & Inconsistency & 1 (0.8\%) & 0 (0.0\%) & 1 (1.0\%) & 0 (0.0\%) & 0 (0.0\%) \\
         & 7. & General Difficulty & 14 (10.8\%) & 7 (5.2\%) & 10 (9.8\%) & 12 (8.5\%) & 15 (11.2\%) \\
        \ \ Personal Values \& Life Fit & 8. & Psychological Resistance & 8 (6.2\%) & 13 (9.7\%) & 6 (5.9\%) & 5 (3.5\%) & 12 (9.0\%) \\
         & 9. & Impacts on/of Lifestyle & 4 (3.1\%) & 22 (16.4\%) & 13 (12.7\%) & 18 (12.7\%) & 8 (6.0\%) \\
        \ \ Accountability & 10. & Uncertainty and Risk & 4 (3.1\%) & 3 (2.2\%) & 8 (7.8\%) & 2 (1.4\%) & 11 (8.2\%) \\
         & 11. & Aversion to AI & 0 (0.0\%) & 1 (0.7\%) & 0 (0.0\%) & 1 (0.7\%) & 0 (0.0\%) \\
        \ \ Others &  &  & 2 (1.5\%) & 4 (3.0\%) & 3 (2.9\%) & 3 (2.1\%) & 5 (3.7\%) \\
        \ \ Total &  &  & 130 & 134 & 102 & 142 & 134 \\
        \midrule
        \multicolumn{3}{l}{\textbf{Willingness to Act (WA)}}\\
        \ \ Effort \& Environmental Constraints & 1. & External Constraints & 9 (7.5\%) & 10 (7.6\%) & 4 (3.7\%) & 5 (3.4\%) & 6 (4.3\%) \\
         & 2. & Degree of Effort & 27 (22.5\%) & 29 (22.0\%) & 37 (34.3\%) & 43 (29.7\%) & 31 (22.5\%) \\
        \ \ Feasibility \& Practicality & 3. & Feasibility Judgment & 24 (20.0\%) & 36 (27.3\%) & 30 (27.8\%) & 40 (27.6\%) & 39 (28.3\%) \\
         & 4. & Gains & 6 (5.0\%) & 14 (10.6\%) & 11 (10.2\%) & 7 (4.8\%) & 11 (8.0\%) \\
        \ \ Personal Values \& Life Fit & 5. & Psychological Resistance & 11 (9.2\%) & 14 (10.6\%) & 4 (3.7\%) & 10 (6.9\%) & 14 (10.1\%) \\
         & 6. & Motives and Values & 12 (10.0\%) & 6 (4.5\%) & 12 (11.1\%) & 14 (9.7\%) & 13 (9.4\%) \\
         & 7. & Unnecessity & 21 (17.5\%) & 10 (7.6\%) & 1 (0.9\%) & 12 (8.3\%) & 10 (7.2\%) \\
        \ \ Accountability & 8. & Uncertainty and Risk & 5 (4.2\%) & 5 (3.8\%) & 2 (1.9\%) & 2 (1.4\%) & 3 (2.2\%) \\
         & 9. & Unfairness & 3 (2.5\%) & 2 (1.5\%) & 1 (0.9\%) & 0 (0.0\%) & 0 (0.0\%) \\
        \ \ Others &  &  & 2 (1.7\%) & 6 (4.5\%) & 6 (5.6\%) & 12 (8.3\%) & 11 (8.0\%) \\
        \ \ Total &  &  & 120 & 132 & 108 & 145 & 138 \\
        \bottomrule
    \end{tabular}
    \caption{Distribution of themes and codes across outcome measures with frequencies and percentages by conditions.}\label{supp_tab:theme_code}
\end{table}

\clearpage
\section{Manipulation Check}\label{supp_sec:manipulation_check}
We performed a two-way ANOVA with recourse-set diversity and size as factors and perceived diversity (Section~\ref{supp_subsec:measure_perceived_diversity}) as an outcome variable to confirm whether these factors effectively alter participants' perceptions of diversity.
As shown in Table~\ref{supp_tab:manipulation_check}, the recourse-set diversity and size significantly changed perceived diversity.
We conclude that the manipulation of perceived diversity via these factors (recourse-set diversity and set size) was successful.
\begin{table}[h]
    \captionsetup{skip=5pt}
    \setlength{\tabcolsep}{0pt}
    \centering
    \small
    \begin{tabular}{lC{0.14\columnwidth}C{0.14\columnwidth}C{0.14\columnwidth}C{0.14\columnwidth}C{0.14\columnwidth}}
        \toprule
        \multicolumn{6}{l}{\textbf{Descriptive statistics}} \\
        \midrule
         & Close--1 & Close--3 & Diverse--3 & Close--7 & Diverse--7 \\
        \midrule
        Mean & --- & 3.276 & 3.714 & 3.804 & 3.873 \\
        $N$  & 139 & 152 & 126 & 168 & 165 \\
        \midrule
        
        \addlinespace[1ex]
        \multicolumn{6}{l}{\textbf{Two-way ANOVA (Diversity $\times$ Size; excluding Close--1)}} \\
        \midrule
         & $SS$ & df$_1$, df$_2$ & $F$ & $p_{\mathrm{adj}}$ & $\eta_p^2$ \\
        \midrule
        Diversity & \ \ 8.485 & 1, 607 & 4.151 & 0.042 & 0.007 \\
        Size & 19.967 & 1, 607 & 9.766 & 0.002 & 0.016 \\
        Diversity $\times$ Size & \ \ 5.127 & 1, 607 & 2.508 & 0.114 & 0.004 \\
        \bottomrule
        
        \multicolumn{6}{l}{$SS$: sum of squares, df: degree of freedom, $F$: F statistics, $p_\mathrm{adj}$: adjusted p-value, $\eta_p^2$: partial eta squared}\\
    \end{tabular}
    \caption{Perceived diversity: descriptive statistics and ANOVA results. }\label{supp_tab:manipulation_check}
\end{table}

\FloatBarrier
\section{Exploratory Analyses with Diagnostic Indicators of Inconsistency and Redundancy}
To further interpret the size-dependent patterns in cognitive load, we conducted exploratory analyses using
diagnostic indicators that capture two undesirable properties of recourse sets: \emph{inconsistency} and
\emph{redundancy}.
We consider these analyses descriptive robustness checks; they do not establish causal mechanisms, and we
leave formal mechanism identification to future work.

\paragraph{Operationalizing inconsistency and redundancy.}
To capture \textbf{inconsistency} across options, we used \emph{sign conflict}, defined as the presence of at
least one pair of options within a set that suggests changing the same feature in opposite directions
(e.g., option A suggests increasing working hours, but option B suggests decreasing working hours).
To capture \textbf{redundancy} among options, we used \emph{common change}, defined as the presence of a
feature that is changed in the same direction by all options within a set (e.g., option A, B, and C in a three-option set suggest increasing the number of side jobs).
Table~\ref{tab:descriptive_indicators} reports the descriptive prevalence of these indicators across conditions.

\paragraph{Model comparison.}
For each outcome $Y$, we compared a base OLS model that includes the factorial design with the diversity--size
interaction,
\begin{equation*}
    \text{M0:}\quad Y \sim \texttt{diversity} \times \texttt{size},
\end{equation*}
against an augmented model that additionally includes a diagnostic indicator $D$,
\begin{equation*}
    \text{M1:}\quad Y \sim \texttt{diversity} \times \texttt{size} + D.
\end{equation*}
As a sensitivity check, we also considered an extended model allowing the association between $D$ and $Y$ to
vary by set size,
\begin{equation*}
    \text{M2:}\quad Y \sim \texttt{diversity} \times \texttt{size} + D + (D \times \texttt{size}).
\end{equation*}
For cognitive-load outcomes (mental demand, effort, and frustration), we used $D$ as the binary presence of
sign conflict (inconsistency).
For negative emotional experience, we used $D$ as the binary presence of common change (redundancy).

\paragraph{Results.}
Table~\ref{supp_tab:model_comparison_indicators} summarizes the model comparisons.
Adding sign conflict (inconsistency) improved model fit for mental demand, whereas it did not improve fit for effort and showed only a suggestive trend for frustration.
Adding common change (redundancy) did not improve model fit for negative emotional experience under this operationalization.
Across outcomes, adding the interaction term $D \times \texttt{size}$ did not yield further improvement, suggesting that these indicators explain only limited portions of the observed size-dependent patterns.
\begin{table}[t]
    \centering
    \setlength{\tabcolsep}{5pt}
    \small
    \begin{tabular}{llccccc}
        \toprule
        Outcome $Y$ & Indicator $D$ & $\Delta$AIC & $\Delta$BIC & $\Delta R^2$ & $p$ (M0 vs M1) & $p$ (M1 vs M2) \\
        \midrule
        Mental demand & Inconsistency: sign conflict (binary) 
            & -5.27 & -0.85 & 0.0115 & 0.0073 & 0.7161 \\
        Effort & Inconsistency: sign conflict (binary) 
            & +2.00 & +6.41 & 0.0000 & 0.9773 & 0.4334 \\
        Frustration & Inconsistency: sign conflict (binary) 
            & -0.74 & +3.67 & 0.0044 & 0.0993 & 0.8115 \\
        Negative emotional experience & Redundancy: common change (binary) 
            & +1.22 & +5.63 & 0.0013 & 0.3790 & 0.4285 \\
        \bottomrule
    \end{tabular}
    \caption{
    Exploratory model comparisons using diagnostic indicators of \textbf{inconsistency} and \textbf{redundancy}.
    We compared a base model (M0: $Y \sim \texttt{diversity} \times \texttt{size}$) against an augmented model adding an indicator $D$
    (M1: $+D$), and, as a sensitivity check, a model additionally including $D \times \texttt{size}$ (M2: $+D+D\times\texttt{size}$).
    Reported $\Delta$AIC/$\Delta$BIC/$\Delta R^2$ are changes from M0 to M1 (negative $\Delta$AIC indicates better fit).
    $p$-values are from nested-model $F$-tests.
    }
    \label{supp_tab:model_comparison_indicators}
\end{table}